\newfont{\mathbb}{bbold12}
\newfont{\Bbf}{bbold8}
\newcommand{\ep}{\epsilon}
\newcommand{\pl}{\partial}
\newcommand{\be}{\begin{equation}}
\newcommand{\ee}{\end{equation}}
\newcommand{\bea}{\begin{eqnarray}}
\newcommand{\eea}{\end{eqnarray}}
\newcommand{\beas}{\begin{eqnarray*}}
\newcommand{\eeas}{\end{eqnarray*}}
\newcommand{\R}{\mathbb R}    

\documentclass[12pt]{JHEP3}
\usepackage{epsfig,amsfonts,mathrsfs,youngtab}
\title{Giant Gravitons on Deformed PP-waves}
\author{Alex Hamilton\\
        Department of Physics\\
	Columbia University, New York\\
	NY 10027 USA\\
	\email{hamilton@phys.columbia.edu}}
\author{Jeff Murugan\\
       	Department of Mathematics and Applied Mathematics\\
	University of Cape Town\\ Private Bag, Rondebosch 7700\\ 
	South Africa\\
	\email{jeff@nassp.uct.ac.za}}

\abstract{The recently constructed Lunin-Maldacena deformation of $AdS_{5}\times S^{5}$ is 
known to support two inequivalent Penrose limits that lead to BPS pp-wave geometries. In this 
note, we construct new giant graviton solutions on these backgrounds. A detailed study of the 
spectra of small fluctuations about these solutions reveals a remarkably rich structure. In 
particular, the giants that we contruct fall into two classes, one of which appears to remain stable 
in the Penrose limit independently of the strength of the deformation. The other class of giants, 
while more difficult to treat analytically, seems to exhibit a shape deformation not unlike the so-
called ``squashed giants" seen in the pp-wave with a constant NS $B$-field turned on. Some 
consideration is also given to the associated giant operators in the BMN limit of the dual ${\cal N}
=1$ SYM gauge theory.} 
\keywords{pp-wave, giant gravitons}
\begin{document}

\section{Introduction}
\label{Introduction}
By now, the gauge theory/gravity correspondence needs little in the way of motivation. Indeed, since its
concrete realization in Maldacena's AdS/CFT conjecture \cite{Maldacena} nearly a decade ago, this remarkable duality has revolutionized the ways in which we think both about quantum field theories and their gravity duals. From an understanding of the emergence of the entropy of a black hole from its constituent microstates \cite{Microstates}, to the emergence of spacetime itself \cite
{Emergence}, the power of the correspondence is surpassed only by its stubborn resistance to direct proof. To date though, not only has the conjecture sucessfully withstood an enormous barrage of tests and checks, it has also emerged as one of the most promising approaches to understanding the strongly coupled behaviour of non-Abelian gauge theories. Indeed, the eventual goal 
of this program is nothing less than a weakly coupled gravitational dual to strongly coupled QCD, one that will be able to shed light on the non-perturbative structure of strong interactions (see, for example, \cite{AdS-QCD} and references therein).\\ 

\noindent
Fuelled largely by the discovery of $D-$brane degrees of freedom in string theory \cite{Polchinski1}, development on the gravity side has been equally impressive. Among this class of extended objects, the giant gravitons of the $AdS_{5}\times S^{5}$ solution of type IIB supergravity \cite{McGreevy,Giant-refs} furnish a particularly rich background in which to study non-perturbative effects in the string theory \cite{Non-pert-string}. As BPS objects blown up on an $S^{3}$ contained in either the $AdS_{5}$ or $5-$sphere parts of the geometry, computations associated to these giant gravitons are protected by powerful non-renormalization theorems and can be extrapolated from weak to strong coupling. Moreover, while giants carry all the same quantum numbers as their energetically degenerate point counterparts, as expanded states in the theory, their interactions are much softer. These interactions however, are difficult to calculate in the supergravity theory. Fortunately, the operators corresponding to giant gravitons have been identified and studied in some detail in the dual ${\cal N}=4$ super Yang-Mills theory with gauge groups $U(N)$ \cite{UN-giants} and $SU(N)$ \cite{SUN-giants}. In the $U(N)$ gauge theory at least, these giant operators are Schur polynomials of the form\footnote{For the special case of giants blown up on the $S^{3}\subset S^{5}$, these Schur polynomials reduce to the more familiar subdeterminant operators identified in the second of refs\cite{UN-giants}.}
\begin{eqnarray}
  \chi_{R}\left(X\right) = \frac{1}{n!}\sum_{\sigma\in S_{n}}\,\chi_{R}(\sigma)\,{\rm tr}\left(\sigma
  X\right)\,,\nonumber
\end{eqnarray}
constructed from the Higgs fields of the SYM multiplet and characterised by Young diagrams of the representation $R$ of $U(N)$. Consequently, the dynamics of the giant graviton is encoded in the Feynman diagrams of the gauge theory via the multi-point correlators of these Schur operators. The study of these operators and the realisation that these states afford a description in terms of free fermions of a single matrix model (see the first ref. of \cite{UN-giants} and \cite{Free-fermion}) has led to a remarkable surge of activity focussed on extracting information about the supergravity background from the gauge theory (see \cite{gauge2SUGRA} and references thereof) culminating in a complete classification of all the classical half-BPS solutions of the type IIB supergravity \cite{LLM}.
This said though, most of what is known about giant gravitons and their dual operators is known in the highly supersymmetric setting of the $AdS_{5}\times S^{5}$/${\cal N}=4$ SYM duality. If we are to 
believe that the duality is an exact one eventually leading to a gravity dual 
of full $4$-dimensional QCD then it is crucial that we understand and extend the gauge/gravity 
dictionary to less supersymmetric backgrounds\footnote{Ideally, we would like to be able to extend 
it further to non-conformal theories as well but we will adopt the time-honored tradition of learning 
to walk before trying to fly.}. \\

\noindent
A significant step toward a systematic study of less supersymmetric systems was taken by Leigh and Strassler \cite{Leigh-Strassler} in their construction of a three parameter family of ${\cal N}=1$ SYM gauge theories obtained by an exactly marginal deformation of the ${\cal N}=4$ superpotential. Then, by exploiting the fact that, at least for a one parameter subset, this Leigh-Strassler deformation may 
be rewritten as a Moyal-like deformation\footnote{For a recent extension of this star product to capture a more general form of the Leigh Strassler deformation see \cite{Bundzik}}, 
\begin{eqnarray}
 \Phi_{i}*\Phi_{j} = e^{i\pi\gamma(Q^{1}_{i}Q^{2}_{j} - Q^{2}_{i}Q^{1}_{j})}\Phi_{i}\Phi_{j},\nonumber
\end{eqnarray}
Lunin and Maldacena showed that the dual supergravity solution \cite{LM} could be constructed from the type IIB $AdS_{5} \times S^{5}$ by a suitable single-parameter deformation of the 5-sphere. This realization has stimulated an enormous resurgence of interest in string theory on this deformed background including, but not limited to, extensions of the Lunin-Maldacena deformation to the eleven dimensional geometries of the Einstein-Sasaki form $AdS_{5}\times Y^{p,q}$ \cite{Einstein-Sasaki}, deformations of  ${\cal N}=1$ and ${\cal N}=2$ theories \cite{Dipole-defs} and a number of studies of semiclassical strings in this background \cite{Deformed-semiclassical} . A particularly insightful contribution to this set of ideas came in \cite{Frolov} where it was shown that the Lunin-Maldacena geometry can equivalently be reached by a sequence of T-dualities and shifts of the angular coordinates by the deformation parameter $\gamma$ - so-called TsT transformations. If, on the other hand, the different angular directions are shifted by {\it different} parameter values, say $\gamma_{i}$, the resulting supergravity background is a completely non-supersymmetric deformation\footnote{The superconformal Lunin-Maldacena deformation then, is a special case in which all three tori are deformed by the same amount, $\hat{\gamma}_{1}=\hat{\gamma}_{2}=\hat{\gamma}_{3}=\hat{\gamma}$} of $AdS_{5}\times S^{5}$ whose gauge theory dual can by systematically constructed from an appropriate generalization of the Leigh-Strassler deformation \cite{FRT}. Remarkably, this multi-parameter deformation is, in some ways, simpler than the single-parameter Lunin-Maldacena deformation. This is particularly true of the study of giant gravitons on these geometries. Generally very difficult to construct, it was rather astutely observed in \cite{DISS} that many of the computations simplify significantly when the deformation in the direction of motion of the $D3-$brane decouples ($\hat{\gamma}_{1} = 0$). Unlike the undeformed geometries previously explored, these giant gravitons are no longer energetically degenerate with the point graviton and consequently unstable, a conclusion supported in the spectrum of small fluctuations about the giant.\\

\noindent
In this article, we revisit the construction of giant graviton solutions in the single-parameter Lunin-Maldacena background. With the deformation coupling to the direction of motion of the $D-$brane, a direct construction of the giant graviton solution in the supergravity proves to be very difficult and, to date, remains an unsolved problem. There is, however, a limit - the Penrose limit - in which the analysis simplifies significantly \cite{BFHP}. This in itself is hardly surprising; afterall the lightcone gauge string action is already known to drastically simplify in this background, enough that the pertubative string spectrum is exactly solvable \cite{BMN}. The Penrose limit of a given geometry is obtained by focussing on a null geodesic in the background. The limiting geometry is the so-called pp-wave . For $AdS_{5}\times S^{5}$, the Penrose limit comes from focussing on a great circle of the $S^{5}$
and then boosting to the infinite momentum frame along this geodesic. There are, of course, 
an infinite number of limits that can be taken, one for each great circle. However, since any great 
circle may be rotated into any other by the $SO(6)$ isometry of the $5$-sphere, there is 
essentially only one resulting pp-wave. On the other hand, the Penrose limit of the deformed 
background is not unique. There are, in fact, two different limits that result in BPS backgrounds. 
One of these is the usual pp-wave background in magnetic coordinates and the other, a 
homogeneous pp-wave \cite{NP,DMSS,Mateos}. We would like to determine 
whether these backgrounds support giant graviton solutions.\\

\noindent
Toward this end, the next section of this article is devoted to a summary of the constuction of giant gravitons on the maximally supersymmetric type IIB pp-wave. We pay particular attention to the different null geodesics about which the Penrose limit can be taken and how they are related. Thereafter, in the interests of self-containment, we discuss the Lunin-Maldacena marginal deformation of $AdS_{5}\times S^{5}$ and the two Penrose limits that result in BPS geometries. In section 4. we show that, indeed both of these deformed pp-waves support giant gravitons exhibiting a remarkably rich structure. In section 5, we initiate an exploration of some of this structure in the spectrum of small fluctuations about the giant graviton equilibrium configuration and conclude, finally, with some comments on the dual gauge theory operators and speculation on future directions.
\section{Giant gravitons, pp-waves and orbits of $S^{5}$.}
\label{pp-giants}
In the context of $AdS_{5}\times S^{5}$ giant gravitons are, by now, well known to be a KK-
mode blown up along a transverse $S^{3}$ contained in the $5-$sphere\footnote{Or, in the case 
of ``dual giants", the $AdS_{5}$} part of the geometry. This blowing up is a consequence of the 
motion of the graviton along some geodesic - a great circle - of the $S^{5}$ through which a 
$5-$form flux is threaded. On the other hand, the pp-wave background is obtained from the $AdS
$ geometry by focussing in on a particular null geodesic, ostensibly by asking what geometry a 
particle boosted to the infinite momentum frame along the geodesic ``sees". It makes sense 
then, to ask if the giants found on the $AdS_{5}\times S^{5}$ background survive the Penrose 
limit. As far as we are aware, this question was first posed and answered in the affirmative in 
\cite{tak-tak}. There, the authors found that giant gravitons in the maximally 
supersymmetric type IIB pp-wave divide into two types \cite{Skenderas-Taylor} depending on the 
geodesic along which the Penrose limit is taken. If the orbit of the graviton coincides with the 
geodesic along which the Penrose limit is taken the resulting giant graviton is a spherical $D3-
$brane whose worldvolume extends along $x^{+}$ and the spatial directions of an $S^{3}$ 
contained in the $5-$sphere. These $D-$branes couple directly to the RR-field and, while they 
don't permit an exact worldsheet description, they do afford an alternate description in terms of 
mutiple $D-$strings whose moduli are argued to be the fluctuations of the giant graviton \cite{DJM}. If, however, the Penrose limit is taken along a geodesic different to the one 
that the $D3$ is moving along, the resulting configuration has Neumann boundary conditions in 
the $x^{+}$ and $x^{-}$ directions as well as two spatial directions (which could be in either the $5-
$sphere or the AdS part of the geometry). A full light-cone gauge worldsheet analysis of 
these $D-$branes reveal that they preserve $16$ supersymmetries but only if they are rotating in 
some transverse two-plane in the pp-wave background. As facinating as these latter solutions are, 
we find the structure of the former (giant graviton) solutions on the deformed pp-wave rich enough 
on its own that, at least for the purposes of this note, we will restrict our attention to them alone. 
A study of the second class of $D-$branes in the Lunin-Maldacena deformed background 
remains an open problem.\\

\noindent
More detailed studies of these configurations, their supersymmetries, fluctuations and 
responses to certain non-vanishing $B-$fields may be found in refs. \cite{sadri-jabbari,prokushkin-jabbari}. In order to facilitate comparison with our later results, we now, very 
briefly but systematically review the construction of giant graviton solutions in the maximally 
supersymmetric type IIB pp-wave background. The construction works equally well when the 
Penrose limit is taken about $(J,0,0),(0,J,0),(0,0,J)$ or $(J,J,J)$ geodesics. In particular, we 
show how the two sets of solutions are related by a light-cone-time-dependent rotation of 
coordinates (or, in the jargon of general relativists, a coordinate transformation between 
Brinkmann and Rosen coordinates). In the undeformed pp-wave background this 
transformation is more-or-less trivial, reflecting the fact that the Penrose limit is essentially unique - 
any orbit on the $5-$sphere may be rotated into any other via the $SO(6)$ isometry group. This, 
we will find, is no longer the case for the pp-waves obtained from the deformed Lunin-Maldacena 
background.\\

\noindent
The maximally supersymmetric type IIB pp-wave\footnote{In this and what follows, we have absorbed the usual mass parameter in the pp-wave into the definition of $x^{-}$}
\begin{eqnarray}
  ds^{2} &=& -2dx^{+}dx^{-} - \left( \sum_{i=1}^{4} (x^{i})^{2} + \sum_{a=5}^{8}(x^{a})^{2}
  \right)(dx^{+})^{2}\nonumber\\
   &+& \sum_{i=1}^{4}(dx^{i})^{2} + \sum_{a=4}^{8}(dx^{a})^{2},
  \label{pp-wave}
\end{eqnarray}
obtains from the $10-$dimensional $AdS_{5}\times S^{5}$ geometry with metric (written in 
global 
coordinates)
\begin{eqnarray}
  ds^{2} = R^{2}\left(-\cosh^{2}\rho\,dt^{2} + d\rho^{2} + \sinh^{2}\rho\,d\Omega_{3}^{2} + \cos^
{2}\alpha\,d\phi_{1}^{2} + d\alpha^{2} + \sin^{2}\alpha\,d\widetilde{\Omega}_{3}^{2}\right),\nonumber
\end{eqnarray}
by rescaling\footnote{Here, $r^{2} = \sum_{i}(x^{i})^{2}$ and $y^{2} = \sum_{a}(x^{a})^{2}$.}
\begin{eqnarray}
  x^{+}=\frac{1}{2}(t+\phi_{1})\,,\qquad x^{-} = R^{2}(t-\phi_{1})\,,\qquad \rho = \frac{r}{R}\,,
\qquad 
\alpha = \frac{y}{R}\,, 
\end{eqnarray}
and sending $R\rightarrow\infty$. Additionally, the only components of the $5-$form flux $F_
{(5)}
$ that survives the limit are those with a plus index,
\begin{eqnarray}
  F_{+1234} = F_{+5768} = {\rm const.}\,,
  \label{Five-form-flux}
\end{eqnarray}
with the constant fully determined once a normalisation of $F_{(5)}$ is specified. 
Physically, this is the geometry near the trajectory - the $(J,0,0)$ orbit - of a particle moving 
with large angular momentum along the $\phi_{1}$ direction and sitting at $\rho = \alpha = 0$. 
When this particle is a $D3-$brane, its action, in lightcone gauge and with a spherical ansatz for 
the brane worldvolume, can be written
\begin{eqnarray}
 S = -\frac{T_{3}\lambda}{g_{s}}\int d\tau d\theta d\phi_{2} d\phi_{3} \left[ r^{3}\sqrt{2\lambda
\nu 
+ \lambda^{2}} - \lambda r^{4}\right]\,,
\label{pp-wave-giant-action}
\end{eqnarray}
where $\lambda$ and $\nu$ are defined through $X^{+}=\lambda \tau$ and $X^{-}=\nu\tau$ respectively.
The corresponding Hamiltonian, as a function of the radius of the $D3$ worldvolume, is 
minimized at $r = 0$, the point graviton, and $r = \sqrt{p^{+}g_{s}/2\pi^{2}T_{3}}$, the giant 
graviton.\\

\noindent
In the above discussion, the $S^{5}$ part of the geometry is parameterized by the five angles $
(\alpha, \theta, \phi_{1}, \phi_{2}, \phi_{3})$ in terms of which the $3-$sphere metric is explicitly 
written as
\begin{eqnarray}
  d\widetilde{\Omega}_{3}^{2} = d\theta^{2} + \cos^{2}\theta\,d\phi_{2}^{2} + \sin^{2}\theta\,d
\phi_
{3}^{2}\,.\nonumber
\end{eqnarray}
There are clearly three $U(1)$ isometries related to translations along each of $\phi_{1}$, $
\phi_{2}$ and $\phi_{3}$. A general orbit in this background is a linear superposition of these three 
isometries. A particle moving on the $(J,J,J)$ orbit then, has equal angular momentum in each 
of these directions. To take a Penrose limit along a null geodesic associated to this orbit, we 
define the new angular variable $\psi = (\phi_{1} + \phi_{2} + \phi_{3})/3$ with periodicity $2\pi$. 
The geodesic we will focus on is given by
\begin{eqnarray}
  t = \psi,\quad \rho = 0,\quad \alpha = \alpha_{0},\quad \theta = \frac{\pi}{4},\quad \psi = \phi_
{1} 
  = \phi_{2} = \phi_{3}\,.
\end{eqnarray}
The geometry in the neighbourhood of this geodesic is recovered by defining lightcone 
coordinates as
\begin{eqnarray}
  x^{+} = \frac{1}{2}(t + \psi)\,,\qquad x^{-} = R^{2}(t - \psi),
\end{eqnarray}
rescaling appropriate coordinates as
\begin{eqnarray}
  \rho = \frac{r}{R},\quad \alpha &=& \alpha_{0} + \frac{y^{1}}{R},\quad \theta = \frac{\pi}{4} + 
\sqrt
  {\frac{3}{2}}\frac{y^{2}}{R},\quad \phi_{1} = \psi - \sqrt{2}\frac{y^{3}}{R},\nonumber\\
  {}\\
  \phi_{2} &=& \psi + \frac{y^{3} - \sqrt{3}y^{4}}{\sqrt{2}R},\quad \phi_{3} = \psi + \frac{y^{3}+
\sqrt
 {3}y^{4}}{\sqrt{2}R},\nonumber
\end{eqnarray}
and taking $R\rightarrow\infty$. The result is, of course, a pp-wave, this time with metric
\footnote{Now, $\tilde{r}^{2} = \sum_{i}(y^{i})^{2}$ and $\tilde{y}^{2} = \sum_{a}(y^{a})^{2}$ and, 
incidentally, we have absorbed the ``mass" parameter $\mu$ into the definition of the lightcone 
variables.}
\begin{eqnarray}
  ds^{2} &=& -2dx^{+}dx^{-} - \tilde{y}^{2}(dx^{+})^{2} + 4\left( y^{1}\,dy^
  {3} + y^{2}\,dy^{4}\right)dx^{+} +d\tilde{y}^{2} + d\tilde{r}^{2}\nonumber\\
  &+& \tilde{y}^{2}\,d\Omega_{3}^{2} + \tilde{r}^{2}\,d\widetilde    
  {\Omega}_{3}^{2}. 
  \label{magnetic-pp-wave}
\end{eqnarray}
The non-vanishing components of the $5-$form in this background remains the same as in eq.
(\ref{Five-form-flux}). String theory on this background was previously studied in \cite{NP,Pando} where the similarity between this metric and that of the motion 
of a particle in a magnetic field earned it the moniker of ``magnetic pp-wave". As expected, all 32 
of the supersymmetries of $AdS_{5}\times S^{5}$ survive the Penrose limit and, since the large $J
$ limit of the dual gauge theory is independent of the choice of $U(1)$ R-charge group, the gauge 
dual of this background is the same truncation of ${\cal N}=4$ SYM as in \cite{BMN}.\\

\noindent
Even though the two backgrounds in eqs.(\ref{pp-wave}) and (\ref{magnetic-pp-wave}) {\it 
seem} very different, they are essentially the same, just written in different coordinates. This is not 
unexpected since this magnetic pp-wave is just the geometry near one particular great circle 
(the $(J,J,J)$ orbit) and any such great circle can be rotated into any other under the action of the 
$SO(6)$ isometry of the $S^{5}$. This rotation may be made more explicit first by making the 
change of variables
\begin{eqnarray}
  x^{+}&\rightarrow& x^{+},\nonumber\\
  x^{-}&\rightarrow& x^{-} + y^{1}y^{3} + y^{2}y^{4},\nonumber
\end{eqnarray}
followed by the lightcone time-dependent coordinate rotation
\begin{eqnarray}
   \left[
     \begin{array}{c}  
       z^{1} \\
       z^{2} \\
       z^{3} \\
       z^{4} \\
     \end{array}
   \right] = 
   \left[
    \begin{array}{cccc}
      \cos x^{+} & 0 & -\sin x^{+} & 0 \\
      0 & \cos x^{+} & 0 & -\sin x^{+} \\
      \sin x^{+} & 0 & \cos x^{+} & 0 \\
      0 & \sin x^{+} & 0 & \cos x^{+}
    \end{array}
   \right] 
    \left[
     \begin{array}{c}  
       \tilde{y}^{1} \\
       \tilde{y}^{2} \\
       \tilde{y}^{3} \\
       \tilde{y}^{4} \\
     \end{array}
   \right].
   \label{Rotation}
\end{eqnarray}  
Implementing this change of variables in the magnetic pp-wave metric gives
\begin{eqnarray}
  ds^{2} &=& -2dx^{+}dx^{-} - \left(\tilde{y}^{2} + z^{2}\right)(dx^{+})^{2} + d\tilde{y}^{2} + \tilde
{y}^
{2}d\Omega_{3}^{2} + dz^{2} + z^{2}d\widetilde{\Omega}_{3}^{2}\,;\nonumber 
\end{eqnarray}
precisely the $(J,0,0)$ metric of (\ref{pp-wave}). The corresponding $5-$form field is
\begin{eqnarray}
  F_{(5)} = \frac{4}{g_{s}}\,dx^{+}\wedge {\Bigl(} dz^{1}\wedge dz^{2}\wedge dz^{3} \wedge dz^
{4}
  + dy^{5}\wedge dy^{6}\wedge dy^{7} \wedge dy^{8}{\Bigr)}\,.\nonumber
\end{eqnarray}
In these coordinates, the construction of giant gravitons in this background is relatively 
straightforward. After all, having made the rotation to the $z$ coordinates, it remains only to
\begin{itemize}
  \item
  Parameterize the $3-$dimensional worldvolume of the giant in static coordinates as
  \begin{eqnarray}
    Z^{1} &=& z\,\cos\theta\,\cos\phi_{2},\nonumber\\
    Z^{2} &=& z\,\sin\theta\,\cos\phi_{3},\nonumber\\
    Z^{3} &=& z\,\cos\theta\,\sin\phi_{2},\nonumber\\
    Z^{4} &=& z\,\sin\theta\,\sin\phi_{3}.\nonumber
  \end{eqnarray}
  \item
  Substitute this ansatz into the worldvolume metric in lightcone gauge and,
  \item
  Compute the corresponding energy functional $E(z)$ together with it's turning points.
 \end{itemize}
This simplicity does not, unfortunately, translate to the $\hat{\gamma}-$deformed pp-wave. 
Nevertheless we can still extract some useful lessons from this exercise. In particular, inverting 
the coordinate transformation (\ref{Rotation}) gives the appropriate coordinatization of the giant 
worldvolume in magnetic coordinates as
\begin{eqnarray}
  \tilde{Y}^{1} &=& z\,\cos\theta\,\cos(\phi_{2}- x^{+}),\nonumber\\
  \tilde{Y}^{2} &=& z\,\sin\theta\,\cos(\phi_{3}- x^{+}),\nonumber\\
  \tilde{Y}^{3} &=& z\,\cos\theta\,\sin(\phi_{2} - x^{+}),\nonumber\\
  \tilde{Y}^{4} &=& z\,\sin\theta\,\sin(\phi_{3} - x^{+}).\nonumber 
\end{eqnarray}
These coordinates are manifestly $x^{+}$, lightcone time dependent. Physically, the giant on this 
pp-wave background also rotates with equal angular momentum in both the $(\tilde{y}^{1},\tilde{y}
^{3})$ and $(\tilde{y}^{2},\tilde{y}^{4})$ planes. It is precisely this form of the magnetic coordinates 
that we will use of the construction of giant gravitons in the deformed pp-wave of \cite{DMSS,Mateos}. 
Finally, to close this section it is worth mentioning that in lightcone coordinates, the $D3-$brane 
action is, not surprisingly, identical to (\ref{pp-wave-giant-action}). 

\section{The Lunin-Maldacena geometry and its Penrose limits}
\label{Lunin-Maldacena}
 In principle, the Lunin-Maldacena construction \cite{LM} is identical to that used in generating the 
holographic duals of noncommutative field theories \cite{MR}. It hinges on the use of an 
$SL(2,\R)$ transformation of the full $SL(3,\R)\times SL(2,\R)$ duality group of type IIB 
supergravity compactified along a corresponding $U(1) \times U(1)$ two-torus. In practice, if the 
metric of this two-torus is denoted by $g$ and the NS-NS $2-$form by $B$ then the Lunin-
Maldacena deformation is implemented by making the replacement
\begin{eqnarray}
  \tau \equiv B + i\sqrt{g} \longrightarrow \tau_{\gamma} = \frac{\tau}{1 + \gamma\tau}\,.
  \label{LM-def}  
\end{eqnarray}
Since this deformation is an exactly marginal one, the AdS factor remains unchanged. In order 
to apply this construction to the gravitational $AdS_{5} \times S^{5}$ background\footnote{Here, $
(\mu_{1},\mu_{2},\mu_{3}) = (\cos\alpha,\sin\alpha\,\cos\theta,\sin\alpha\,\sin\theta)$ and the 
common radius of the $AdS_{5}$ and $S^{5}$ parts of the geometry, $R^{4}= 4\pi e^{\phi_{0}}Nl_
{s}^{4}$. Note also that  in the original undeformed geometry, the NS-NS $2-$form $B = 0$. and the coordinates have been chosen to manifestonly a  $U(1)^{3}$ subgroup of the full $SO(6)$ isometry of the round $5-$sphere.},
\begin{eqnarray}
  ds^{2} &=& R^{2} \left( -\cosh^{2}\rho\,dt^{2} + d\rho^{2} + \sinh^{2}\rho\,d\Omega_{3}^{2} + 
\sum_{i=1}^{3} d\mu_{i}^{2} + \mu_{i}^{2}\,d\phi_{i}^{2}\right)\,,\nonumber\\
  F_{(5)} &=& 4R^{4} e^{-\phi_{0}}\left( \omega_{AdS_{5}} + \omega_{S^{5}}\right)\,,\\
  e^{\phi} &=& e^{\phi_{0}}\,,\nonumber
\end{eqnarray}
we first define three new angles through 
\begin{eqnarray}
  \phi_{1} = \psi + \varphi_{1} + \varphi_{2},\quad \phi_{2} = \psi - \varphi_{1},\quad \phi_{3} = \psi - 
\varphi_{2}\,,
\end{eqnarray}
then choose a $U(1) \times U(1)$ subgroup that acts by shifting $\varphi_{1}$ and $\varphi_{2}$. 
The corresponding two-torus along which we compactify has $\tau$ parameter
\begin{eqnarray}
  \tau = iR\sqrt{\mu_{1}^{2}\mu_{2}^{2} + \mu_{1}^{2}\mu_{3}^{2} 
  + \mu_{2}^{2}\mu_{3}^{2}}\,,\nonumber
\end{eqnarray}
Finally, applying the Lunin-Maldacena deformation (\ref{LM-def}) gives the type IIB supergravity multiplet,
\begin{eqnarray}
  ds^{2} &=& R^{2}{\Bigl[} - \cosh^{2}\rho\,dt^{2} + d\rho^{2} + \sinh^{2}\rho\,d\Omega_{3}^{2} + 
  d\alpha^{2} + G\cos^{2}\alpha\,d\phi_{1}^{2} + \sin^{2}\alpha {\bigl(} d\theta^{2}\nonumber\\
  &+& G\cos^{2}\theta\,d\phi_{2}^{2} + G\sin^{2}\theta\,d\phi_{3}^{2}{\bigr)} + \hat{\gamma}^{2} 
  G\cos^{2}\alpha \sin^{4}\alpha \cos^{2}\theta \sin^{2}\theta \left( d\phi_{1} + d\phi_{2} + d\phi_{3}
  \right)^{2}{\Bigr]},\nonumber\\
  B_{(2)} &=& \hat{\gamma}R^{2} G {\Bigl(} \sin^{2}\alpha \cos^{2}\alpha \cos^{2}\theta\,d\phi_{1}
  \wedge d\phi_{2} + \sin^{2}\alpha \cos^{2}\alpha \sin^{2}\theta\, d\phi_{3}\wedge d\phi_{1}
  \nonumber\\
  &+& \sin^{4}\alpha \cos^{2}\theta \sin^{2}\theta\, d\phi_{2}\wedge d\phi_{3}{\Bigr)},\nonumber\\
  F_{(3)} &=& -\frac{4\hat{\gamma}}{g_{s}}R^{2}\cos^{2}\alpha \sin^{3}\alpha \cos\theta \sin\theta
  \,d\alpha \wedge d\theta \wedge \left( d\phi_{1} + d\phi_{2} + d\phi_{3}\right),\nonumber\\
  F_{(5)} &=& \frac{4}{g_{5}}R^{4}\left( \cosh\rho \sinh^{3}\rho\, dt\wedge d\rho \wedge d
  \Omega_{3} + G \cos\alpha \sin^{3}\alpha d\phi_{1}\wedge d\alpha \wedge d\widetilde{\Omega}_{3}\right),
\nonumber 
\end{eqnarray}
where $ G^{-1} = 1 + \hat{\gamma}^{2} \left( \cos^{2}\alpha \sin^{2}\alpha + \sin^{4}\alpha \cos^
{2}\theta \sin^{2}\theta \right)$ and the rescaled $\hat{\gamma} = R^{2}\gamma$. Since the deformation is continuous, this one-parameter family of backgrounds are topologically identical to that of the original round sphere. Clearly also, only the sphere part of the geometry suffers any deformation with the original $SO(6)$ isometry group breaking to a $U(1)^{3}$. This has important consequences for the Penrose limits of this background that we wish to consider. In the undeformed case, all null geodesics lying inside the $5-$sphere can be rotated into each other by the action of the $SO(6)$ isometry group. Now, with only a $U(1)^{3}$ group remaining, this is no longer the case. Indeed there are two distinct classes of BPS geodesics whose Penrose limits result in non diffeomorphic pp-wave metrics. With $\tau$ as a worldline coordinate and setting $t=\phi_{1}=\phi_{2}=\phi_{3}=\tau$, the maximum circles on the deformed $S^{5}$ with
\begin{eqnarray}
 (\mu_{1}^{2},\mu_{2}^{2},\mu_{3}^{2}) = (1,0,0),(0,1,0),(0,0,1),\nonumber
\end{eqnarray}
furnish one set and the null geodesics with
\begin{eqnarray}
   (\mu_{1}^{2},\mu_{2}^{2},\mu_{3}^{2}) = (1/3,1/3,1/3),\nonumber
\end{eqnarray}
the other. Since each of the $(J,0,0),(0,J,0)$ and $(0,0,J)$ orbits may be rotated into each other with the residual isometry after the $SL(2,\R)$ action on the $5$-sphere, we need only focus on one of this set of orbits. To take the Penrose limit about the $(J,0,0)$ orbit say, on $S_{\gamma}^{5}$, we set
\begin{eqnarray}
 \rho &=& \frac{y}{R},\qquad\qquad\quad\, \alpha = \frac{r}{R},\nonumber\\
 t &=& x^{+} + \frac{x^{-}}{2R},\qquad \phi_{1} = x^{+} - \frac{x^{-}}{2R},\nonumber
\end{eqnarray}
and then take $R\rightarrow\infty$. With a little algebra, it is easily seen that the resulting 
background fields are
\begin{eqnarray}
  ds^{2} &=& -2dx^{+}dx^{-} - \left( y^{2} + \left( 1 + \hat{\gamma}^{2}\right)r^{2}\right)(dx^{+})^
{2} + dy^{2} + y^{2}\,d\Omega^{2}_{3} + dr^{2} + r^{2}\,d\widetilde{\Omega}_{3}^{2},  \nonumber\\
  B_{(2)} &=& \hat{\gamma}r^{2}\left( \cos^{2}\theta\,dx^{+}\wedge d\phi_{2} + \sin^{2}\theta\,d
\phi_{3}\wedge dx^{+}\right),\nonumber\\
  F_{(5)} &=& \frac{4}{g_{s}}\left( y^{3}\, dx^{+}\wedge dy\wedge d\Omega + r^{3}\,dx^{+}\wedge 
dr \wedge d\widetilde{\Omega}_{3}\right)\nonumber\\
&{}&\\
  C_{(4)} &=& -\frac{1}{g_{s}}\left( y^{4}\,dx^{+}\wedge d\Omega_{3} + r^{4}\, dx^{+}\wedge d\widetilde{\Omega}_{3}\right)\nonumber\\
  C_{(2)} &=& 0,\nonumber\\
  F_{(3)} &=& dC_{(2)} = 0,\nonumber 
  \label{J00-multiplet}
\end{eqnarray}
with $d\widetilde{\Omega}_{3} = \sin\theta \cos\theta\,d\theta\wedge d\phi_{1}\wedge d\phi_{3}$ the standard volume element on the round $3-$sphere. Closed strings on this pp-wave were quantized, and their relation to the BMN limit of the dual ${\cal N}=1$ field theory was studied initially in \cite{NP} and then in the current context in \cite{LM}. In a remarkable feat of reverse engineering, the authors of \cite{NP} found that, in lightcone gauge $x^{+}=\tau$, the theory on the worldsheet is a massive one with a transverse oscillation spectrum in the deformed sphere directions of
\begin{eqnarray}
  \omega_{n} = \sqrt{1+\left( \frac{n}{|p^{+}|}\pm\hat{\gamma}\right)^{2}}\,,
  \label{J00-spectrum}
\end{eqnarray}  
by working backwards from knowing the spectrum of anomalous dimensions of the corresponding gauge theory operators. This result was later confirmed by direct computation in \cite{LM}.\\

\noindent
An altogether different Penrose limit can be taken by focussing on states having charges near $(J,J,J)$. These live near the null geodesic $\tau = \psi$ with $\alpha_{0}=\cos^{-1}(1/\sqrt{3})$ and $\theta_{0}=\pi/4$. Setting
\begin{eqnarray}
 \theta = \frac{\pi}{4} + \sqrt{\frac{2}{3}}\frac{x^{1}}{R}\,,\quad \alpha &=& \alpha_{0} - \frac{x^{2}}{R}\,,\quad \rho = \frac{y}{R}\,,\nonumber\\
 \varphi_{1} = \frac{\tilde{x}^{3}}{R}\,\quad \varphi_{2} = \frac{\tilde{x}^{4}}{R}\,,\quad t &=& x^{+} + \frac{x^{-}}{2R}\,,\quad \psi = \frac{x^{-}}{2R} - x^{+}\,,\nonumber
\end{eqnarray}
redefining
\begin{eqnarray}
 x^{3} = \sqrt{\frac{2}{3+\hat{\gamma}^{2}}}\left( \tilde{x}^{3} + \frac{1}{2}\tilde{x}^{4}\right)\,,\quad
 x^{4} = \sqrt{\frac{3}{2(3+\hat{\gamma}^{2})}}\tilde{x}^{4}\,,\nonumber
\end{eqnarray}
and taking the $R\rightarrow\infty$ limit gives the pp-wave metric
\begin{eqnarray}
  ds^{2} &=& -2dx^{+}dx^{-} - \left( \sum_{a=5}^{8}(x^{a})^{2} + \frac{4\hat{\gamma}^{2}}{3+\hat{\gamma}^{2}}\left((x^{1})^{2} + (x^{2})^2{}\right)\right)(dx^{+})^{2} + \sum_{a=5}^{8}(dx^{a})^{2}\nonumber\\
  &+& \sum_{i=1}^{4}\,(dx^{i})^{2} + \frac{4\sqrt{3}}{\sqrt{3+ \hat{\gamma}^{2}}}(x^{1}dx^{3}
  +x^{2}dx^{4})dx^{+}\,.
  \label{JJJ-def-pp-wave}
\end{eqnarray}
In this same limit, the remaining fields in the IIB multiplet are
\begin{eqnarray}
  B_{(2)} &=& \frac{\hat{\gamma}}{\sqrt{3}}\,dx^{3}\wedge dx^{4} + \frac{2\hat{\gamma}}{\sqrt{3+\hat{\gamma}^{2}}}\,dx^{+}\wedge\left( x^{1}dx^{4} - x^{2}dx^{3}\right)\,,\nonumber\\
  C_{(2)} &=& \frac{2\hat{\gamma}}{\sqrt{3}g_{s}}\,dx^{+}\wedge \left( x^{2}dx^{1} - x^{1}dx^{2} \right)\,,\nonumber\\
  &{}&\\
  F_{(5)} &=& \frac{4}{g_{s}}\,dx^{+}\wedge\left( dx^{1}\wedge dx^{2}\wedge dx^{3}\wedge dx^{4} +
  dx^{5}\wedge dx^{6}\wedge dx^{7}\wedge dx^{8} \right)\,,\nonumber\\
 e^{2\phi} &=& \frac{1}{1+\hat{\gamma}^{2}}e^{2\phi_{0}}\,.\nonumber
 \label{JJJ-multiplet}
\end{eqnarray}
It is immediatly evident from the non-vanishing RR $2-$form and constant dilaton that this pp-wave differs rather nontrivially from the former pp-wave. Certainly, since the $3-$brane couples to the RR $2-$form, it might already be expected that the physics of giant gravitons is markedly more sophisticated on this background. Closed strings in this background, their supersymmetries and dual gauge theory operators were first studied in \cite{DMSS,Mateos}. There, it was noticed independently that a change of coordinates from $x^{-}$ to $x^{-}+\sqrt{3/(3+\hat{\gamma}^{2})}(x^{1}x^{3} + x^{2}x^{4})$ brings the $(J,J,J)$ pp-wave metric into the homogeneous plane wave form \cite{Homogeneous-pp}
\begin{eqnarray}
  ds^{2} &=& -2dx^{+}dx^{-} - \left[ \sum_{a=5}^{8}(x^{a})^{2} + \frac{4\hat{\gamma}^{2}}{3 + \hat{\gamma}^{2}} \left( (x^{1})^{2} + (x^{2})^{2}\right)\right] (dx^{+})^{2}\nonumber\\
   &+& \sum_{i=1}^{8}(dx^{i})^{2} + 2\sqrt{\frac{3}{3 + \hat{\gamma}^{2}}}\left( x^{1}dx^{3} - 
   x^{3}dx^{1} + x^{2}dx^{4} - x^{4}dx^{2}\right)dx^{+}\,,
   \label{Homogeneous}
\end{eqnarray} 
after which it is a short, if somewhat technical, task to quantize the string sigma model and compute the full closed string spectrum which, in lightcone gauge and units of $2\pi\alpha' = 1$ reads
\begin{eqnarray}
  \omega_{n} = 1 \pm \sqrt{1 + 4n^{2}}\,.
  \label{JJJ-spectrum}
\end{eqnarray} 
Crucially, the closed string spectrum is completely independent of the deformation. Evidently the deformation of the geometry is exactly compensated for by the now non-vanishing $B-$field. Added support for this conclusion is given by a computation of the spectrum of anomalous dimensions of the corresponding operators in the dual ${\cal N}=1$ gauge theory. The operators in question were proposed in \cite{DMSS} to be of the form
\begin{eqnarray}
  {\cal O}_{(p)}^{\hat{\gamma}} = \sum_{n,m}\,{\rm tr}\left( {\cal O}_{n,m}^{\hat{\gamma}}\right)\delta_{m-n,p}\,,
  \label{Deformed operators}
\end{eqnarray}
where $p=0,1,...,J$ and the ${\cal O}_{n,m}^{\hat{\gamma}}$ are built out of the complex Higgs's $\Phi^{1},\Phi^{2}$ and $\Phi^{3}$ in the gauge theory by taking one of the $\Phi$'s to define a ``background" lattice and inserting the other two as impurities hopped to the $n-$ and $m-$th postions respectively. The matrix of anomalous dimensions of the ${\cal O}_{(p)}^{\hat{\gamma}}$ was then numerically checked to be identical, in the large $J$ limit, to that of the corresponding undeformed operators \cite{DMSS}. This is in precise agreement with the string theory prediction.\\

\noindent
One final point that deserves mentioning about this particular pp-wave geometry concerns its supersymmetry content. Our starting point, the type IIB $AdS_{5}\times S^{5}$ background is maximally supersymmetric  in $10-$dimensions. It's pp-wave limit is, famously, also maximally supersymmetric and preserves all 32 of the supersymmetries. The Lunin-Maldacena $\hat{\gamma}-$deformation, of course, breaks a number of these. Exactly how many, can be computed by solving the Killing spinor equations for the backgrounds. Fortunately this rather technical exercise can be circumvented by using the fact that the $\hat{\gamma}-$deformation and Penrose limits commute \cite{LM,Mateos}. In other words, starting from $AdS_{5}\times S^{5}$, Penrose limits may be taken along $(J,0,0)$ and $(J,J,J)$ orbits to get the respective pp-waves discussed in the previous section. An approriate set of coordinates is then selected and compactified on a torus and the $SL(2,\R)$ action applied along this torus. Finally, decompactifying the torus directions gives the $(J,0,0)$ and $(J,J,J)$ deformed pp-waves respectively. Of the 32 Killing spinors associated to the original pp-wave backgrounds, only those that are independent of the torus coordinates will survive the $SL(2,\R)$ transformation.  For the ($J,0,0$) deformed pp-wave, these include the standard 16 supersymmetries generically preserved by pp-waves plus an additional 8 so-called supernumenary supersymmetries giving a total of 24 altogether \cite{NP}. On the other hand, while the $(J,J,J)$ deformed pp-wave preserves the same standard 16 Killing spinors, only 4 of the remaining 16 survive the $SL(2,\R)$ transformation as supernumenary supersymmetries. This gives a total of 20 for this background.  

\section{Giants on the deformed pp-wave}
\label{deformed giants}
Having developed the background on which we will work, we now proceed to the main focus of this work, namely the existence and stability of giant gravitons on these geometries. We begin with the deformed $(J,0,0)$ pp-wave of \cite{LM,NP} and then proceed to the more difficult case of the pp-wave constructed around the $J_{\phi_{1}} = J_{\phi_{2}} = J_{\phi_{3}} $ geodesic. The physics of the problem here consists of a $D3$-brane wrapping the $S^{3}$ in the deformed $5$-sphere and orbiting with fixed angular momentum along some great circle on $S_{\gamma}^{5}$. The Penrose limit in both cases is {\it along} this orbital direction - the $\phi_{1}$ equatorial circle in the former case and the $\psi$ orbit in the latter. Consequently, the worldvolume of the giant is of spherical topology (a $(+,3,0)$ brane in the notation of \cite{Skenderas-Taylor}. Had we boosted along a different direction, the brane worldvolume would have included both lightcone directions $x^{+}$ and $x^{-}$ and two transverse spatial directions. We do not know whether these $D-$brane configurations are supported on this background and leave this for future consideration.\\

\subsection{$(J,0,0),(0,J,0)$ and $(0,0,J)$ orbits}
On this class of geodesics, it will suffice to consider a $D3-$brane moving along the $\phi_{1}$ equitorial circle with fixed angular momentum and then take the Penrose limit along this same orbit. Consequently, we will take as an ansatz for the giant graviton solution,
\begin{eqnarray}
 \phi_{1} = \omega t;\quad \sigma^{1} = \theta;\quad \sigma^{2} = \phi_{2};\quad \sigma^{3} = 
\phi_
{3}.
\end{eqnarray}
with static gauge specified by
\begin{eqnarray}
 X^{+} &=& \lambda \tau;\qquad\qquad\qquad X^{-} = \mu\tau;\nonumber\\
 X^{1} &=& r\cos\theta \cos\phi_{2};\qquad X^{2} = r\cos\theta\sin\phi_{2};\\
 X^{3} &=& r\sin\theta \cos\phi_{3};\qquad X^{4} = r\sin\theta\sin\phi_{3}.\nonumber
\end{eqnarray}
The lightcone gauge constraint $X^{+} = \tau$ is enforced by evaluating all expressions at $\lambda=1$. Clearly, from the form of the deformed metric, any excitations in the AdS direction are unchanged by the deformation. Consequently, the background also supports AdS giants ({\it i.e.} those that are blown up along the $S^{3}$ of the $AdS_{5}$ part of the geometry). While these AdS giant solutions themselves 
are not affected by the deformation, their spectra of small fluctuations will be. however, we leave the study of these solutions for future work. For now, let's turn off any AdS fields;
\begin{eqnarray}
  X^{a} = 0\,\qquad a = 5,6,7,8.\nonumber
\end{eqnarray}
In this pp-wave limit, since the dilaton remains constant up to an overall factor of $G$ and $G
\rightarrow 1$, the $D3$-brane action in this geometry is simply
\begin{eqnarray}
  S = -\frac{T_{3}}{g_{s}}\int\,d^{4}\sigma\,\, \sqrt{-{\rm det}\,({\rm P}[g + B])} + T_{3}\int\,{\rm P}\, 
  [C_{(4)}]\,
\end{eqnarray}
where, as usual, ${\rm P}[X]$ denotes the pullback of the spacetime field $X$ to the brane 
worldvolume and $g_{s} = \exp(\phi_{0})$ is the $10$-dimensional string coupling. In what 
follows, it will prove more useful to rewrite the NS $B$-field in cartesian coordinates on the brane 
worldvolume. This is easily seen to be
\begin{eqnarray}
  B_{(2)} = \hat{\gamma}\,dX^{+}\wedge(X^{1}\,dX^{2} - X^{2}\,dX^{1} + X^{4}\,dX^{3} - X^{3}\,
  dX^{4}).\nonumber 
\end{eqnarray}
Focussing first on the Born-Infeld part of the action, we first define the pullback
\begin{eqnarray}
  D_{\mu\nu} = {\rm P}[g + B]_{\mu\nu} = X^{\alpha}_{\mu}X^{\beta}_{\nu}g_{\alpha\beta} + X^
  {\alpha}_{\mu}X^{\beta}_{\nu}B_{\alpha\beta}, 
\end{eqnarray}
where $\alpha,\beta,...$ are spacetime indices and $\mu\nu,...$ worldvolume ones. With a little 
tedious algebra, the components of $D_{\mu\nu}$ are readily computed as
\begin{eqnarray}
  D_{\mu\nu} =    
  \left[\begin{array}{cccc} 
    -2\lambda\mu - \lambda^{2}(1+\hat{\gamma}^{2})r^{2} & 0 & \hat{\gamma}\lambda r^{2}
    \cos^{2}\theta & - \hat{\gamma}\lambda r^{2}\sin^{2}\theta\\
     0 & r^{2} & 0 & 0 \\
     -\hat{\gamma}\lambda r^{2}\cos^{2}\theta & 0 & r^{2}\cos^{2}\theta & 0\\
     \hat{\gamma}\lambda r^{2}\sin^{2}\theta & 0 & 0 & r^{2}\sin^{2}\theta   
     \end{array}\right].\nonumber
\end{eqnarray}
Notice that all off-diagonal contributions to $D_{\mu\nu}$ come from the deformation. 
Consequently, turning off $\hat{\gamma}$ manifestly reduces $D_{\mu\nu}$ to that of the 
undeformed pp-wave giant graviton. Computing the determinant of this matrix, we find that
\begin{eqnarray}
  {\rm det}\,D_{\mu\nu} = - r^{6}\sin^{2}\theta\,\cos^{2}\theta (2\lambda\mu + \lambda^{2}r^{2}).
\end{eqnarray}
On substituting this into the Born-Infeld action, the corresponding Lagrangian is read off as
\begin{eqnarray}
  L_{DBI} = -\frac{T_{3}\widetilde{\Omega}_{3}}{g_{s}}r^{3}\sqrt{2\lambda\mu + \lambda^{2}r^{2}},
\end{eqnarray}
where $\widetilde{\Omega}_{3} = \int\,d\theta d\phi_{2} d\phi_{3}\,\sin\theta \cos\theta = 2\pi^{2}$ 
is the volume of the transverse $3$-sphere that the giant expands into. That the volume element 
that arises in the DBI action is the standard one on the round sphere is a good consistency check of 
our sperical ansatz. We will return to this point a little later in the construction of giant gravitons 
moving along $(J,J,J)$ orbits. Returning to our present construction, it remains to compute the Chern-Simons contribution to the full action. Since the only RR field that survives this Penrose limit is 
$C_{(4)}$, this is particularly easy
\begin{eqnarray}
  S_{CS} &=& T_{3}\int\,C_{(4)} = \frac{T_{3}}{g_{s}}\int\,r^{4}\sin\theta\cos\theta\,dX^{+}\wedge 
  d\theta\wedge d\phi_{2}\wedge d\phi_{3}\nonumber\\
  &=& \frac{T_{3}\lambda}{g_{s}}\int d\tau\int d\theta d\phi_{2} d\phi_{3}\,\, r^{4}\,
  \sin\theta\cos\theta.\nonumber
\end{eqnarray}
The corresponding Chern-Simons Lagrangian is 
\begin{eqnarray}
  L_{CS} = M\lambda r^{4},
\end{eqnarray}
where we have taken the liberty of defining $M \equiv T_{3}\widetilde{\Omega}_{3}/g_{s}$. 
Putting this together then, gives the full $3$-brane Lagrangian
\begin{eqnarray}
  L = - M\left[ r^{3}\sqrt{2\lambda\mu + \lambda^{2}r^{2}} - \lambda r^{4}\right].
  \label{J00-Lagrangian}
\end{eqnarray}
Notice that this Lagrangian is {\it independent of the deformation parameter}. This is not unlike 
what happens to closed strings quantized on the pp-wave of \cite{DMSS, Mateos}. Recall that there, 
the Lunin-Maldacena deformation had the effect of turning on an NS $B$-field as well as deforming 
the geometry and both these contributions exactly cancel so that the closed string spectrum is $
\hat{\gamma}-$independent. Of course, the deformation affects more than just the NS sector of the 
type IIB supergravity; it also turns on various form fields of the RR sector. Since strings couple only to the geometry and the $B-$field none of these form fields mattered in the study of the closed string spectrum. The $3-$brane, on the other hand, also couples to the $2-$ and $4-$form fields $C_{(2)}$ and $C_{(4)}$ respectively. However, while both $C_{(2)}$  and $C_{(4)}$ are affected by the deformation, it is only 
$C_{(4)}$ that survives the $(J,0,0)$ Penrose limit. It is hardly surprising then, that the 
Lagrangian (\ref{J00-Lagrangian}) is independent of the deformation parameter.\\

\noindent
Returning to the construction of the giant graviton, the lightcone momentum of the $3-$brane is
\begin{eqnarray}
  p^{+} = -\frac{\delta L}{\delta \mu} = \frac{M\lambda r^{3}}{\sqrt{2\lambda\mu + \lambda^{2}r^{2}}}  
  \,,
  \label{J00-momentum}
\end{eqnarray}
with an associated lightcone Hamiltonian 
\begin{eqnarray}
  p^{-} = H_{lc} = -\frac{\delta L}{\delta \lambda} = \frac{Mr^{3}(\mu + \lambda r^{2})}{\sqrt{2
\lambda\mu + \lambda^{2}r^{2}}} - Mr^{4}\,.\nonumber
\end{eqnarray} 
This can be rewritten in terms of the lightcone momentum by solving the expression for the 
latter for $\mu$ and substituting into that for $p^{-}$. Consequently,
\begin{eqnarray}
  H_{lc} = \frac{M^{2}}{2p^{+}}r^{6} + \frac{p^{+}}{2}r^{2} - Mr^{4}\,.
\end{eqnarray} 
As a function of the $3-$brane radius, this Hamiltonian is extremized at
\begin{eqnarray}
  r = \left\{
       \begin{array}{c}
         0\\
         \sqrt{\frac{p^{+}}{3M}}\\
         \sqrt{\frac{p^{+}}{M}}\\
       \end{array} \right.  \,. 
\end{eqnarray}
As in the undeformed pp-wave, $r=0$ is the point graviton while $R_{G}\equiv \sqrt{p^{+}/M}$ 
is the radius of the giant graviton. A plot of the lightcone Hamiltonian, $H_{lc}$, in units of $M = p^{+}=1$ is given in Figure 1. 
\begin{figure}[hp]
    {\epsfig{file=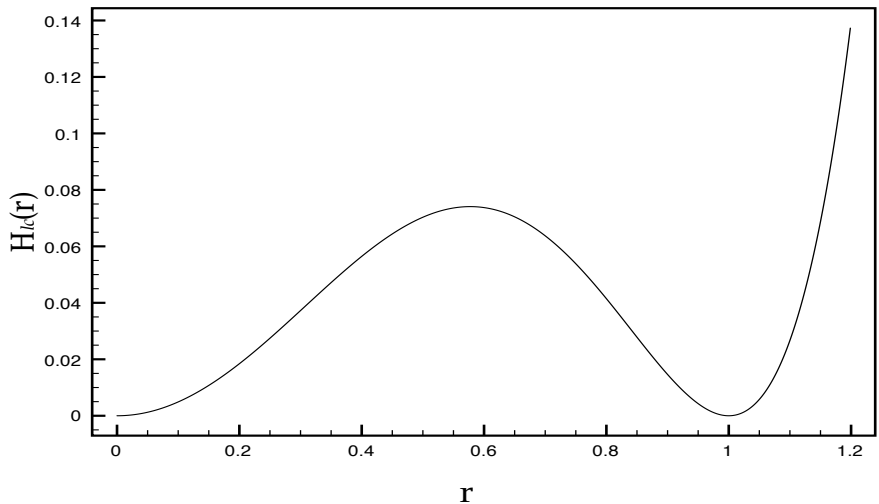,width=12cm,height=7cm}
    \caption{The giant graviton lightcone Hamiltonian on the $(J,0,0)$ pp-wave exhibits two 
    degenerate minima corresponding to the point and giant gravitons respectively.}}
\end{figure}  
\noindent
With no way to energetically distinguish  between the point and giant gravitons on this background, we expect that the giant solution should be stable. This expectation does, however, need to be verified in the spectrum of small fluctuations about the giant. We do so in the next section. 

\subsection{$(J,J,J)$ orbits}
The second of the two Penrose limits that result in BPS backgrounds - specifically that taken about the $(J,J,J)$ orbit - is a little more difficult to treat.  The IIB mutiplet in this background is given by $(3.6)$ and $(3.7)$. The construction of giant gravitons in this background will be facilitated by choosing a slightly different gauge in which $x^{-}\rightarrow x^{-} + \sqrt{\frac{3}{3+\hat{\gamma}^{2}}}(x^{1}x^{3} + x^{2}x^{4})$. With this change of coordinates,
\begin{eqnarray}
  ds^2 &=& -2 dx^{+} dx^{-} - \left( \sum_{a=5}^{8}(x^{a})^2 + 4\frac{\hat{\gamma}^{2}}{3+\hat{\gamma^{2}}}\sum_{i=1}^{2}(x^{i})^{2} \right) (dX^+)^2 + \sum_{a=5}^{8}(dx^{a})^{2}\nonumber\\
   &+& \sum_{i=1}^{4} (dx^{i})^{2}
   + 2 \sqrt{\frac{3}{3+\hat{\gamma}^{2}}} \left(x^{1} dx^3 - x^{3} dx^{1} + x^{2} dx^{4} - x^{4} dx^{2}\right)\,,\nonumber\\
  B_{(2)} &=& \sqrt{\frac{\hat{\gamma}^{2}}{3}} dx^{3} \wedge dx^{4} + 2 \sqrt{\frac{\hat{\gamma}^{2}}{3+\hat{\gamma}^{2}}} (X_1 dX_4 - X_2 dX_3)\,, \nonumber\\
  &{}&\\
  C_{(4)} &+& C_{(2)} \wedge B_{(2)} = \frac{1}{3!} \frac{(3+\hat{\gamma}^2)}{3g_s}\epsilon_{ijkl}x^{i} dx^{j} \wedge dx^{k} \wedge dx^{l}\,, \nonumber\\
  e^{\phi} &=& \frac{3}{3+\hat{\gamma}^{2}} g_s \,.\nonumber
\end{eqnarray}
Note, in particular that the RR $2-$form is no longer vanishing and so its coupling to the $D3-$brane via the $C_{(2)}\wedge B_{(2)}$ Chern-Simons term is no longer trivial. Since the extended structure of the giant graviton is essentially supported by the RR flux, any change in the flux is likely to change the size and/or shape of the giant. Similar studies of pp-wave giant gravitons were carried out in \cite{sadri-jabbari,prokushkin-jabbari} where it was reported that when a constant transverse $B-$field is turned on in the background, giant gravitons respond by a deformation of their shape. The situation in this background is markedly different though.  Here, in addition to a non-vanishing $C_{(2)}$, the NS $B-$field induced by the $\hat{\gamma}-$deformation is certainly not constant. Proceeding with the construction then, our ansatz for the giant graviton on this background is
\begin{eqnarray}
  \psi= \omega t\,,\quad \sigma^{1}=\theta\,,\quad\sigma^{2}=\phi_{2}\,\quad \sigma^{3}=\phi_{3}\,.
  \label{JJJ-ansatz}
\end{eqnarray}
Now, as promised, building on the lessons learnt from the undeformed pp-wave in magnetic coordinates, we parameterize the $D3-$brane worldvolume by
\begin{eqnarray}
  X^{+} &=& \lambda \tau\,,\qquad\qquad\qquad\qquad\qquad\qquad\quad\!\!
  X^{-}  = \mu \tau\,; \nonumber\\
  X^{1} &=& r \cos\theta\,\cos \left(\phi_{2} - \sqrt{\frac{3}{3+\hat{\gamma}^{2}}} X^{+}\right), \,\,
  X^{2} = r \sin\theta\, \cos \left(\phi_{3} -\sqrt{\frac{3}{3+\hat{\gamma}^{2}}} X^{+}\right)\,; \nonumber\\
  X^{3} &=& r \cos\theta\,\sin \left(\phi_{2} -\sqrt{\frac{3}{3+\hat{\gamma}^{2}}} X^{+}\right),\,\,
  X^{4} = r \sin\theta\,\sin \left(\phi_{3} - \sqrt{\frac{3}{3+\hat{\gamma}^{2}}} X^{+}\right)\,. \nonumber
\end{eqnarray}
Note that, as in the undeformed pp-wave, this parameterization of the giant worldvolume is manifestly time-dependent. As in that case, we could perform a coordinate rotation to static Rosen coordinates but this comes at a cost of a significantly more cumbersome action for the $D3-$brane. Instead, we choose to work in the Brinkmann coordinates above, where at least the physics of the problem is slightly more transparent. As before, we denote the pullback of the combination of the metric and NS $2-$form by $D_{\mu \nu} = {\rm P}[g + B_{(2)}]_{\mu\nu}$. The time components of $D_{\mu\nu}$ are then  easily computed as
\begin{eqnarray}
  D_{\tau \tau}&=& -2 \lambda \mu - \lambda^{2} \frac{3}{3+\hat{\gamma}^{2}} \left( r^{2} 
  + 4\frac{\hat{\gamma}}{3}(\cos^{2} \theta\cos^{2} \phi_{2} 
  + \sin^{2} \theta \cos^{2} \phi_{3}) \right)\,,\nonumber\\
  D_{\tau \theta} &=& -D_{\theta\tau} = 
  \lambda \sqrt{\frac{\hat{\gamma}^{2}}{3+\hat{\gamma}^{2}}}
  r^{2} \left(\cos^{2} \theta \cos \phi_{2} \sin\phi_{3} 
  + \sin^{2}\theta \sin\phi_{2} \cos\phi_{3}\right)\,,\nonumber\\
  D_{\tau\phi_{2}} &=& -D_{\phi_{2}\tau} = 
  - \lambda  \sqrt{\frac{\hat{\gamma}^{2}}{3+\hat{\gamma}^{2}}}r^2 \cos\theta \sin\theta \cos\phi_{2} 
  \cos\phi_{3}\,,\nonumber\\
  D_{\tau\phi_{3}} &=& -D_{\phi_{3}\tau} = 
  \lambda \sqrt{\frac{\hat{\gamma}^{2}}{3+\hat{\gamma}^{2}}} r^{2} \cos\theta \sin\theta \cos\phi_{2} 
  \cos\phi_{3}\,,\nonumber
\end{eqnarray}
while its spatial components are
\begin{eqnarray}  
D_{rs} = \left[ 
 \begin{array}{ccc}
    r^{2} & - \frac{\hat{\gamma}}{\sqrt{3}} r^{2} \cos^{2} \theta \cos\phi_{2} \sin\phi_{3} & 
    -  \frac{\hat{\gamma}}{\sqrt{3}} r^{2} \sin^{2}\theta \sin\phi_{2} \cos\phi_{3} \\
    \frac{\hat{\gamma}}{\sqrt{3}} r^{2} \cos^{2} \theta \cos\phi_{2} \sin\phi_{3} 
    & r^{2} \cos^{2}\theta & \frac{\hat{\gamma}}{\sqrt{3}} r^{2} \cos\theta \sin\theta \cos\phi_{2} 
    \cos\phi_{3} \\
    \frac{\hat{\gamma}}{\sqrt{3}} r^{2} \sin^{2}\theta \sin\phi_{2} \cos\phi_{3} 
    & - \frac{\hat{\gamma}}{\sqrt{3}} r^{2} \cos \theta \sin \theta \cos \phi_2 \cos \phi_3 
    & r^{2} \sin^{2}\theta
 \end{array} \right] \,,\nonumber
\end{eqnarray}
where $r,s = \theta,\phi_{2},\phi_{3}$. For ease of notation, we have set $\tau=0$ above. The time dependence of the solution can be restored by replacing $\phi_{i}$ with $\phi_{i} -\sqrt{\frac{3}{3+\hat{\gamma}^{2}}} X^+$. Then, with $Y^{i}\equiv X^{i}/r$, {\it i.e.} the embedding coordinates of a unit $3-$sphere and a little algebra, the determinant of the pullback can be expressed as 
\begin{eqnarray}
  -\det D_{\mu \nu} &=& r^{6} \cos^{2}\theta \sin^{2}\theta \left[ 2 \lambda \mu \left(1 
  + \frac{\hat{\gamma}^{2}}{3} \left((Y^{1})^2 + (Y^{2})^{2}\right) \right) \right. \nonumber\\
  &+& \left. \lambda^{2} r^{2} \frac{3}{3+\hat{\gamma}^{2}} \left( 1 + 2 \frac{\hat{\gamma}^{2}}{3} 
  \left((Y^{1})^{2} + (Y^{2})^{2}\right) \right)^2 
  + \lambda^2 \frac{\hat{\gamma}^{2}}{3+\hat{\gamma}^{2}} r^2 \left(Y^{1} Y^{4}
   - Y^{2} Y^{3}\right)^{2} \right] \,.\nonumber
\end{eqnarray}
The manifest dependence of this determinant on the $Y^{i}$ - compared to, say the corresponding expression in the $(J,0,0)$ pp-wave - is strong evidence that the giant in this background no longer retains its round $3-$sphere shape.  Evidently, our ansatz for the giant is unable to capture the full shape deformation of the giant and requires some refinment. This is a nontrivial exercise best left for future work. Nevertheless, we can still make some progress - albeit limited - when the Lunin-Maldacena deformation is small.  Toward this end, we will take $\hat{\gamma}$ small and expand all $\hat{\gamma}-$dependent terms to lowest nontrivial order in $\hat{\gamma}$. Expanding to ${\cal O}(\hat{\gamma}^{2})$ then, the $D3-$brane Lagrangian
\begin{eqnarray}
L = -T_3 e^{-\phi} \int \, d^3 \sigma\, \sqrt{-\det D_{\mu \nu}} + T_{3} \int \, {\rm P}\left[C_{(4)} + C_{(2)} \wedge B_{(2)}\right]\,, \nonumber
\end{eqnarray}
becomes
\begin{eqnarray}
  L  &=& - \frac{T_{3}}{g_{s}} \left(1 + \frac{\hat{\gamma}^{2}}{3}\right) r^{3} \sqrt{2 \lambda \mu + \lambda^{2} 
  r^{2}} \int \, d\theta d\phi_{2} d\phi_{3}\, \cos \theta \sin \theta \nonumber\\ 
  & \times & \sqrt{ 1+ \frac{\hat{\gamma}^{2}}{3(2 \lambda \mu + \lambda^{2} r^{2})} \left[ (2 
  \lambda \mu 
  + 4 \lambda^{2} r^{2})\left((Y^{1})^{2} + (Y^{2})^{2}\right) + 
  \lambda^{2} r^{2} \left(Y^{1} Y^{4} - Y^{2} Y^{3}\right)^{2} - \lambda^{2} r^{2} \right] } \nonumber\\
  &+& 2 \pi^{2}\lambda \frac{T_{3}}{g_{s}} \left(1 + \frac{\hat{\gamma}^{2}}{3}\right)r^{4} + 
  \mathcal{O}\left(\hat{\gamma}^{4}\right)\,,\nonumber
\end{eqnarray}
or, using $M=2\pi^{2}T_{3} g_{s}^{-1}$ and the integrals
\begin{eqnarray}
  \int \, d\theta d \phi_{2} d \phi_{3}\, \cos \theta \sin \theta \left((Y^{1})^{2} + (Y^{2})^{2}\right) = \pi^{2}\,,\nonumber
\end{eqnarray}
and
\begin{eqnarray}
   \int \, d\theta d \phi_{2} d \phi_{3}\, \cos \theta \sin \theta 
   \left(Y^{1} Y^{4} - Y^{2} Y^{3}\right)^{2} = \frac{\pi^{2}}{6} \,,\nonumber
\end{eqnarray}
\begin{eqnarray}  
  L= \lambda M r^{4} - M r^{3} \sqrt{2 \lambda \mu + \lambda^{2} r^{2}} + 
  \frac{\hat{\gamma}^{2}}{3} \left[ \lambda M r^{4}  - M r^{3} \frac{\frac{5}{2} \lambda \mu 
  + \frac{37}{24} \lambda^{2} r^{2}}{\sqrt{2 \lambda \mu + \lambda^{2} r^{2}}} \right] 
  + \mathcal{O} \left(\hat{\gamma}^{4}\right)\,.\nonumber
\end{eqnarray}
The corresponding lightcone momentum 
\begin{eqnarray}
  p^{+} &=& -p_{-} = -\frac{\delta L}{\delta \mu} \nonumber\\
  &=& \frac{\lambda M r^{3}}{\sqrt{2 \lambda \mu + \lambda^{2} r^{2}}} + 
  \frac{\hat{\gamma}^{2}}{3} \frac{\lambda M r^{3}}{(2 \lambda \mu
   + \lambda^{2} r^{2})^{3/2}} \left[ \frac{5}{2} \lambda \mu + \frac{23}{24} \lambda^{2} r^{2}\right] 
   + \mathcal{O} \left(\hat{\gamma}^{4}\right) \,.
   \label{JJJ-def-momentum}
\end{eqnarray}
To compute the lightcone Hamiltonian, we need to eliminate any explicit $\mu-$dependence in the expression for $p^{-}$. In anticipation, solving (\ref{JJJ-def-momentum}) for $\mu$ to order
$\hat{\gamma}^{2}$ we get,
\begin{eqnarray}
  \lambda \mu \sim \frac{1}{2} \left( \frac{\lambda M r^{3}}{p^{+}} \right)^{2} 
  - \frac{1}{2} \lambda^{2} r^{2} + \frac{\hat{\gamma}^{2}}{3} 
  \left( \frac{5}{4} \left( \frac{\lambda M r^{3}}{p^{+}} \right)^{2} - \frac{7}{24} \lambda^{2} r^{2} 
  \right)\,. 
\end{eqnarray}
Finally, to lowest nontrivial order in $\hat{\gamma}$, the lightcone Hamiltonian
\begin{eqnarray}
 \!\!\!p^{-} \sim \frac{r^{2}}{2 p^{+}} \left[ M^{2} r^{4} \left( 1 + \frac{5}{6} \hat{\gamma}^{2} \right) 
  - 2 M p^{+} r^{2} \left(1 + \frac{\hat{\gamma}^{2}}{3}\right) + (p^{+})^{2} \left(1 
  + \frac{7}{36} \hat{\gamma}^{2} \right) \right].
  \label{def-JJJ-hamiltonian}
\end{eqnarray}
As a rough check, note that this reduces to the correct undeformed Hamiltonian in the
 $\hat{\gamma} \to 0$ limit.  In fact, writing it as
 \begin{eqnarray}
   H_{lc}\sim \left[\frac{M^{2}}{2p^{+}}r^{6} + \frac{p^{+}}{2}r^{2} - Mr^{4}\right] 
   + \hat{\gamma}^{2}\left[\frac{5M^{2}r^{6}}{12p^{+}} - \frac{Mr^{4}}{3} 
   + \frac{7(p^{+})^{2}r^{2}}{72}\right]\,,
   \nonumber
 \end{eqnarray}
 it is clear that, at least to ${\cal O}(\hat{\gamma}^{2})$, the Hamiltonian is positive semi-definite, vanishing only at $r=0$. We have checked numerically that this property persists even up to order 
 $\hat{\gamma}^{4}$. Figure.2 is a plot of the lightcone Hamiltonian (\ref{def-JJJ-hamiltonian}) as a function of $r$. 
\begin{center} 
\begin{figure}[h]
    {\epsfig{file=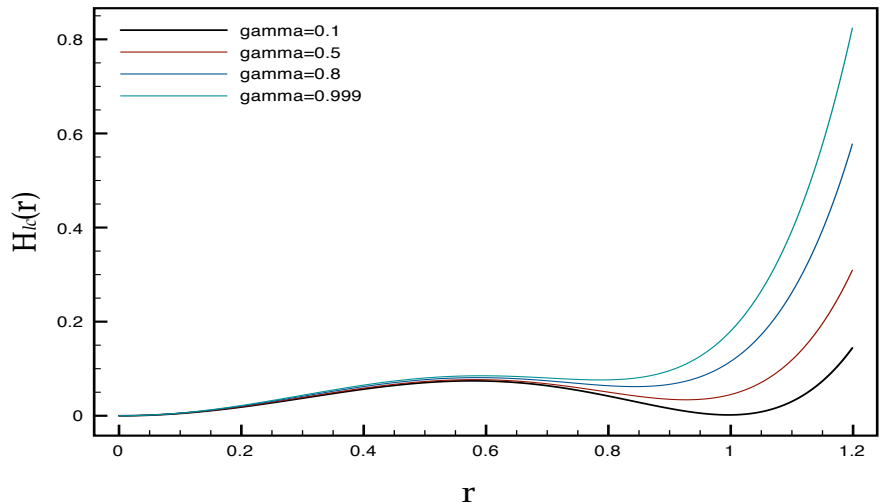,width=12cm,height=7cm}
    \caption{The deformation lifts the degenerate minima in the $(J,J,J)$ pp-wave.}}
\end{figure}
\end{center}
Notice now that the once degenerate vacuum at finite radius is lifted by the deformation. Consequently, the energy of the giant is greater than that of the point graviton on this background and the giant graviton is no longer a stable state in the theory. Indeed, beyond a certain critical value of the deformation parameter $\hat{\gamma}\sim 1$ it seems the second minimum vanishes and the giant is no longer a solution even. This is in good agreement with the results of \cite{prokushkin-jabbari} and, especially \cite{DISS}. Here too, as the deformation is increased from zero, the $C_{(2)}\wedge B_{(2)}$ contribution to the Chern-Simons term grows faster than the four form flux term and eventually dominates. However it is important to bear in mind that our analysis is carried out perturbatively in $\hat{\gamma}$ and any conclusions drawn at large $\hat{\gamma}$ need to be done with care. So, although numerical results up to ${\cal O}(\hat{\gamma}^{4})$ seem to support the reported behaviour of the $D3-$brane lightcone Hamiltonian, there is probably not too much to be gained from estimating the critical value of 
$\hat{\gamma}$ at which the second minimum vanishes - at least until a more detailed study of the deformation of the giant on this pp-wave is completed.
 
\section{Small fluctuations and stability of giants}
The spectrum of small fluctuations about a giant graviton background encodes many interesting properties of the equilibrium configuration \cite{DJM}. For instance, since zero energy modes in the vibration spectrum correspond to a family of solutions each having the same energy and angular momentum, the study of these modes is important in determining the ability of the giant ansatz in capturing all the classical BPS states. More relevantly for us, any perturbative instability in the giant configuration will appear in the fluctuation spectrum as a tachyonic mode. In this section we will systematically investigate the spectrum of small fluctuations about the equilibrium configurations of the pp-wave giant gravitons beginning first with the giant on the maximally supersymmetric type IIB pp-wave and then proceeding to the $\hat{\gamma}-$deformed $(J,0,0)$ background. While the vibration spectrum of giants on the undeformed pp-wave is not unknown \cite{sadri-jabbari, prokushkin-jabbari}, we will rederive it here using the Lagrangian methods of \cite{DJM} primarily by way of review and to establish our notation but also to facilitate comparison with the deformed giant spectrum. Essentially the Hamiltonian computation proceeds by fixing all the conserved quantities in the problem, diagonalising the Hamiltonian\footnote{For small perturbations the Hamiltonian is a quadratic form in the coordinates and momenta representing the perturbation and may be diagonalised using standard methods for the diagonalisation of quadratic forms.} and extracting the classical oscillation frequencies. The problem is that any mode for which the conserved quantities are not held fixed will necessarily be excluded from the spectrum. We will therefore proceed with the Lagrangian analysis of \cite{DJM}.
\subsection{Undeformed giants}
Our starting point here is the giant graviton configuration on the pp-wave (\ref{pp-wave}) supported by the $5-$form flux (\ref{Five-form-flux}). This equilibrium configuration is specified by  
\begin{eqnarray}
 X^{+} &=& \lambda\tau,\quad\qquad\qquad\,
 X^{-} = \nu\tau,\nonumber\\
 X^{1} &=& r_{0}\cos\theta_{1}\cos\theta_{2},\quad
 X^{2} = r_{0}\sin\theta_{1}\cos\theta_{3},\\
 X^{3} &=& r_{0}\cos\theta_{1}\sin\theta_{2},\quad
 X^{4} =r_{0}\sin\theta_{1}\sin\theta_{3}.\,\nonumber   
\end{eqnarray} 
Correspondingly, the perturbed configuration has coordinates\footnote{Without loss of generality, $\lambda$ can be set to unity.}
\begin{eqnarray}
X^{+}&=& \tau, \quad\qquad\quad\,
X^{-} = \nu \tau + \ep \delta x^{-}, \nonumber \\
&{}&\\
r &=& r_0 + \ep \delta r,\quad
X^{a} = \ep \delta x^{a},\nonumber
\end{eqnarray}
and $\sigma^{i} = \theta_{i} \in S^{3}$. This ansatz is then substituted into the $D3-$brane action
\begin{eqnarray}
S=-\frac{T_{3}}{g_{s}} \int d\tau d^{3} \sigma \sqrt{-\det ({\rm P}[g])} + T_{3} \int {\rm P}[C_{(4)}],\nonumber
\end{eqnarray}
and, as we are interested primarily in the fluctuation spectrum, expanded up to quadratic order in the perturbation parameter, $\ep$.  It is then quite straightforward to compute the induced metric on the brane worldvolume. With $D_{\mu\nu}\equiv {\rm P}[g]_{\mu\nu}$, we find
\begin{eqnarray}
  D_{0i} &=& D_{i0} \approx -2 \epsilon \pl_{i} \delta x^{-} + \ep^{2} 
  \left[ (\pl_{\tau} \delta r)(\pl_{i} \delta r) + (\pl_{\tau} \delta x^{a})(\pl_{i} \delta x^{a}) 
  \right], \nonumber \\
  D_{ij} &\approx& r_{0}^{2} g_{ij} + 2\ep r_{0} \delta r g_{ij} + \ep^{2} 
  \left[\delta r^{2} g_{ij} + (\pl_{i} \delta r)(\pl_j \delta r)
   + (\pl_{i} \delta x^{a})(\pl_{j} \delta x^{a}) \right], \nonumber
\end{eqnarray}
and
\begin{eqnarray}
  D_{00}&\approx& -(2\nu + \mu^{2} r_{0}^{2}) 
  \left[1 + \ep \left(\frac{2\pl_{\tau} \delta x^{-} + 2\mu^{2} r_{0} 
  \delta r}{2\nu + \mu^{2}r_{0}^{2}} \right)  \right. \nonumber\\
  &+&\left. \ep^{2} \left( \frac{\mu^{2} \delta r^{2} + \mu^{2} (\delta x^{a})^{2} 
  - (\pl_{\tau} \delta r)^{2} - (\pl_{\tau} \delta x^{a})^{2}}{2\nu 
  + \mu^{2} r_{0}^{2}} \right) \right]. \nonumber
\end{eqnarray}
A judicious application of Stokes' theorem on the $D3-$brane worldvolume allows for the Chern-Simons contribution to the action to easily be written down as
\begin{eqnarray}
 S_{CS} = \frac{T_{3}}{g_{s}} \mu (r_{0} + \ep \delta r)^{4} \int\, d\tau\, d^{3}\sigma\, \sqrt{|g_{ij}|}\,,
\nonumber
\end{eqnarray}
Substituting this, together with the metric pullback terms above, into the worldvolume action, we can read off the term linear in $\ep$ as
\begin{eqnarray}
  \!\!\!\!\!\!\!\!\!\delta S = \ep\frac{T_{3}\Omega_{3}}{g_{s}}\int d\tau 
  \frac{r_{0}^{2}}{\sqrt{2\nu + \mu^{2} r_{0}^{2}}} 
  \left[ \left(6\nu + 4\mu^{2} r_{0}^{2} - 4\mu r_{0} \sqrt{2\nu + \mu^{2} r_{0}^{2}} \right) \delta r 
  + r_{0} \frac{\pl \delta x^{-}}{\pl \tau} \right]\!\!
\label{Linear-fluctuation}
\end{eqnarray}
Since we vary with fixed boundary conditions ({\it i.e.} $\pl_\tau \delta x^-$ vanishes at the endpoints of $\tau$) and the coefficient of $\pl_\tau \delta x^-$ is constant, this term gives no contribution to the first order variation of the action. In order for the radial perturbation to vanish, on the other hand, we must have $\nu=0$. The other value of $\nu$ for which the $\delta r$ coefficient vanishes, $\nu=-\frac{4}{9} \mu^{2} r_{0}^{2}$, is a nonphysical solution and consequently discarded.
Next we evaluate the perturbations second order in $\ep$. With the choice of $\nu=0$ above, these terms assemble as
\begin{eqnarray}
  \delta^{2} S &=& -\ep^{2}\frac{T_{3}}{g_{s}} \mu r_{0}^{3} \int d\tau d^{3}\sigma 
  \sqrt{|g_{ij}|} \frac{\delta r}{2\mu^{2} r_{0}^{2}} 
  \left[ \frac{\pl^{2}}{\pl \tau^{2}} - \mu^{2} \frac{1}{\sqrt{|g_{kl}|}} \pl_{i} 
  \left(\sqrt {|g_{kl}|} g^{ij} \pl_{j} \right) \right] \delta r \nonumber \\
  &-& 2\frac{\delta x^{-}}{2\mu^{2} r_{0}^{3}} \frac{\pl \delta r}{\pl \tau} + 
  2\frac{\delta r}{2\mu^{2} r_{0}^{3}} 
  \frac{\pl\delta x^{-}}{\pl \tau} \nonumber\\
  &+&\frac{\delta x^{-}}{2\mu^{4} r_{0}^{4}} \left[\frac{\pl^{2}}{\pl\tau^{2}} - 
  \mu^{2} \frac{1}{\sqrt{|g_{kl}|}} \pl_{i} \left(\sqrt{|g_{kl}|} g^{ij} \pl_{j} \right) \right]
  \delta x^{-} \nonumber \\
  &+& \frac{\delta x^{a}}{2\mu^2 r_{0}^{2}} \left[ \mu^{2} + \frac{\pl^{2}}{\pl\tau^{2}} 
  - \mu^{2} \frac{1}{\sqrt{|g_{kl}|}} \pl_{i} (\sqrt{|g_{kl}|} g^{ij} \pl_{j} ) \right] \delta x^{a}\,.
  \label{Second-order-flucts}
\end{eqnarray}
In order to analyze the resulting vibration spectrum, the fluctuations are first decomposed into their respective Fourier components:
\begin{eqnarray}
\delta x^{\mu} = \widetilde{\delta x}{}^{\mu}\, e^{-i\omega \tau}\, Y_{l} (\theta_1, \theta_2, \theta_3) \,.
\end{eqnarray}
Here, the $Y_l$ are 4-dimensional spherical harmonics satisfying
\begin{eqnarray}
 \frac{1}{\sqrt{|g_{mn}|}} \pl_{i} \left(\sqrt{|g_{mn}|} g^{ij} \pl_{j} \right) Y_{l} = -Q_{l} Y_{l}\,,
 \label{Spherical-harmonics}
\end{eqnarray}
with the eigenvalues $Q_l = l(l+2)$. Substituting the Fourier expansion into 
(\ref{Second-order-flucts}) we see that the fluctuations in the $X^{a}$ directions are separable. Their spectrum is easily determined to be
\begin{eqnarray}
\omega_{a}^{2} = \mu^{2} (1+Q_{l}) \,.
\end{eqnarray}
The radial and $X^-$ fluctuations are a bit more tricky. They are coupled, and satisfy the second order system 
\begin{eqnarray}
\left[ \begin{array}{cc}
	Q_{l}- \frac{\omega^{2}}{\mu^{2}} & \frac{-2i}{\mu^{2}r_{0}} \omega \\
	\frac{2i}{\mu^{2}r_{0}} \omega &  \frac{Q_{l}}{\mu^{2}r_{0}^{0}} 
	- \frac{\omega^{2}}{\mu^{4}r_{0}^{2}}
\end{array} \right]
\left[ \begin{array}{c}
	\widetilde{\delta r} \\
	\widetilde{\delta x}{}^{-}
\end{array} \right] = 0\,.
\end{eqnarray}
Solving this for $\omega$ leads to the spectrum
\begin{eqnarray}
  \omega_{\pm}^{2} = \mu^{2}\left( 2 + Q_{l} \pm 2\sqrt{1+Q_{l}}\right)\,,
\label{pp-wave-spectrum}
\end{eqnarray}
At this point some comments are in order. Most obvious in the above spectrum is its independence of the radius $r_{0}$ of the brane. This should be compared to the corresponding spectrum of sphere giants on $AdS_{5}\times S^{5}$ where it was demonstrated in \cite{DJM} that a suitable rescaling of the fluctuation fields completely scales out all $r_{0}$ dependence of the small fluctuation action. Secondly, note that as there are no complex frequencies, there are no tachyonic modes. Consequently, these giant gravitons on the pp-wave are stable. This is in perfect agreement with 
the results of \cite{sadri-jabbari,tak-tak}. Moreover, there is a zero mode when $l=0, Q_l=0$ on the negative branch of (\ref{pp-wave-spectrum}). This corresponds to the statement that increasing the size of the giant (which relates to increasing the light-cone momentum) costs no energy, just as we expect.

\subsection{Deformed giants}
Moving on to the deformed case now, we focus on the giant gravitons on the $(J,0,0)$ pp-wave. 
As in the undeformed case above, we take a static, spherical ansatz for the giant, wrapping the deformed $S^3$ in the light cone gauge and allow fluctuations in the transverse directions. The ansatz for the perturbed giant is, as before
\begin{eqnarray}
  X^{+} &=& \tau\,, \qquad\qquad\quad\,\,\,
  X^{-} =  \nu \tau + \epsilon \delta x^{-}\,,\nonumber\\
  X^{1} &=& r \cos \theta \cos \phi_{2}\,, \quad
  X^{2} = r \cos \theta \sin \phi_{2}\,, \\
  X^{3} &=& r \sin \theta \cos \phi_{3}\,, \quad
  X^{4} = r \sin \theta \sin \phi_{3}\,, \nonumber
\end{eqnarray}
where $r = r_{0} + \epsilon \delta r$ and the AdS fluctuations are $X^{a} = \epsilon \delta x^{a}$. Expanded up to second order in $\ep$ the components of the pullback $D_{\mu \nu} = 
{\rm P}[g + B_{(2)}]_{\mu\nu}$ appearing in the Born-Infeld contribution to the $D3-$brane action,
\begin{eqnarray}
  S = -\frac{T_{3}}{g_{s}} \int d\tau \ d\Omega \sqrt{-\det D_{\mu \nu}} 
  + T_3 \int {\rm P}[C_{(4)} + C_{(2)} \wedge B_{(2)}]
  \label{Def-D3-fluctuation-action}
\end{eqnarray}
are
\begin{eqnarray}
  D_{\tau\tau} &=& -2 \nu  - \left(1 + \hat{\gamma}^{2}\right) r_{0}^{2} 
  + \epsilon \left[ -2 \partial_{\tau} \delta x^{-} 
  - 2 \left(1+\hat{\gamma}^{2}\right) r_{0} \delta r\right]\nonumber\\
  &+& \epsilon^{2} \left[-\sum_{a} \left(\delta x^{a}\right)^{2} 
  - \left(1+\hat{\gamma}^{2}\right) \delta r^{2} 
  + \sum_{I = a+r} \left(\partial_{\tau} \delta x^{I}\right)^{2} \right]\,, \nonumber\\
  D_{\tau\theta} &=& D_{\theta\tau} = -\epsilon \partial_{\theta} \delta x^{-} 
  + \epsilon^{2} \sum_{I=a+r} \left(\partial_{\tau} \delta x^{I}\right)
  \left(\partial_{\theta} \delta x^{I}\right)\,, \nonumber\\
  D_{\tau\phi_{2}/\phi_{2}\tau} &=& \pm \hat{\gamma}  r_{0}^{2} \cos^{2} \theta 
  + \epsilon \left[-\partial_{2} \delta x^{-} \pm 2 \hat{\gamma} r_{0} \cos^{2}\theta \delta r\right]
   \nonumber \\
  &+& \epsilon^{2} \left[\sum_{I=a+r} \left(\partial_{\tau} \delta x^{I}\right)
  \left(\partial_{2} \delta x^{I}\right) 
  \pm \hat{\gamma} \cos^{2} \theta \delta r^{2} \right]\,, \\
  D_{\tau\phi_{3}/\phi_{3}\tau} &=&  \mp \hat{\gamma}  r_{0}^{2} \sin^{2} \theta
  + \epsilon \left[ -\partial_{\phi_{3}} \delta x^{-} \mp 2 \hat{\gamma} r_{0} 
  \sin^{2} \theta \delta r\right] \nonumber\\
  &+& \epsilon^{2} \left[ \sum_{I=a+r} \left(\partial_{\tau} \delta x^{I}\right)
  \left(\partial_{\phi_{3}} \delta x^{I}\right) \mp \hat
  {\gamma} \sin^{2} \theta \delta r^{2} \right]\,,\nonumber \\
  D_{ij} &=& r_{0}^{2} g_{rs} + 2 \epsilon r_{0} g_{ij} \delta r + \epsilon^{2} 
  \left[g_{ij} \delta r^{2} + \sum_{I=a+r} \left(\partial_{i} \delta x^{I}\right)
  \left(\partial_{j} \delta x^{I}\right) \right] \,. \nonumber
\end{eqnarray}
Here, $\sum_{I=a+r}$ is shorthand for a summation over both the AdS fluctuations, $\delta x^{a}$ and radial perturbation $\delta r$ and the undeformed spatial metric, $g_{rs}$ is
\[g_{rs} = \left[ \begin{array}{ccc}
	r^2 & & \\
	 & r^2 \cos^2 \theta & \\
	 & & r^2 \sin^2 \theta
\end{array} \right] \,.\]
Then, keeping terms second order and lower in $\epsilon$, the determinant of $D_{\mu\nu}$ is
\begin{eqnarray}
  -\det D_{\mu \nu} 
  &=& r_{0}^{6} \cos^{2} \theta \sin^{2} \theta \left(2 \nu + r_{0}^{2}\right) 
  + \epsilon \left(r_{0}^{5} \cos^{2}\theta \sin^{2} \theta\right) 
  \left[2 r_{0} \partial_{\tau} \delta x^{-} +  \left( 12 \nu + 8 r_{0}^{2}\right) 
  \delta r\frac{}{}\right] \nonumber\\
  &+& \epsilon^{2} \left(r_{0}^{4} \cos^{2} \theta \sin^{2} \theta\right) 
  \left[ \left(30 \nu + 28r_{0}^{2}\right) \delta r^{2} 
  + r_{0}^{2} \sum_{a} \left(\delta x^{a}\right)^{2} \right. \nonumber\\
  &+&  \sum_{I=a+r} \left( \left(2\nu + r_{0}^{2}\right) \sum_{s} \ 
  g^{ii} \left(\partial_{i} \delta x^{I}\right)^{2} + \hat{\gamma}^{2} r_{0}^{2} 
  \left(\partial_{\phi_{2}} \delta x^{I} - \partial_{\phi_{3}} \delta x^{I}\right)^{2} 
  - r_{0}^{2} \left(\partial_{\tau} \delta x^{I}\right)^{2} \right) \nonumber\\
  &+& 12 \left. r_{0} \delta r \left(\partial_{\tau} \delta x^{-}\right) 
  + \sum_{s} g^{ii} \left(\partial_{i} \delta x^{-}\right)^{2} \right] \,.
\end{eqnarray}
Notice that the deformation has cancelled in the zeroth and first order fluctuations.  
This is just a manifestation, at the level of fluctuations, of the NS $B-$field exactly compensating for changes in the metric induced by the Lunin-Maldacena deformation.  Interestingly though, this is not carried through to the higher order terms. At ${\cal O}(\epsilon^{2})$ for example, the 
fluctuation spectrum exhibits a ``magnetic field" like splitting due to the deformation. The Chern-Simons term on the other hand is trivially integrated over the sphere to give
\begin{eqnarray}
  S_{CS} = \frac{T_{3}}{g_{s}} \Omega_{3} \int \ d\tau 
  \left(r_{0}^{4} + 4 \epsilon r_{0}^{3} \delta r + 6 \epsilon^2 r_{0}^{2} 
  \delta r^{2}\right)\,,
\end{eqnarray}
to second order in $\epsilon$.  Putting this all together, defining $m\equiv T_{3}/g_{s}$ and expanding the square root gives
\begin{eqnarray}
  S &=& -m \Omega_{3} r_{0}^{3} \int d\tau \ \sqrt{2 \nu + r_{0}^{2}} - r_{0}\nonumber \\
  &-& \epsilon \frac{m r_{0}^{3}}{\sqrt{2 \nu + r_{0}^{2}}} \int  d\Omega_{3} d\tau \,
   \left[ \partial_\tau \delta x^{-} + 
  \frac{\delta r}{r_{0}} \left(6 \nu + 4r_{0}^{2} - 4r_{0} \sqrt{2 \nu + r_{0}^{2}} \right) \right] \nonumber\\
  &-& \epsilon^{2} \frac{mr_{0}}{2 \sqrt{2 \nu + r_{0}^{2}}} \int d\Omega_{3} d\tau \  
  \left[ \left( 30 \nu +  28r_{0}^{2} 
  - \frac{(6 \nu + 4r_{0}^{2})^{2}}{2 \nu + r_{0}^{2}} 
  - 12r_{0} \sqrt{2 \nu + r_{0}^{2}} \right) \delta r^{2} \right. \nonumber\\
  &+&  \sum_{I=a+r} \, \left( (2\nu + r_{0}^{2}) \sum_{i} \ g^{ii} \left(\partial_{i} \delta x^{I}\right)^{2} 
  + \hat{\gamma}^{2} r_{0}^{2} \left(\partial_{\phi_{2}} \delta x^{I} 
  - \partial_{\phi_{3}} \delta x^{I}\right)^{2} 
  - r_{0}^{2} \left(\partial_{\tau} \delta x^{I}\right)^2 \right)\nonumber\\
  &+& r_{0}^{2} \sum_{a} \ \left(\delta x^{a}\right)^{2} 
  + \left. 2 \left( \frac{6 \nu + 2r_{0}^{2}}{2 \nu + r_{0}^{2}} \right)r_{0} \delta r \left(\partial_{\tau} 
  \delta x^{-}\right) + \sum_{i} g^{ii} \left(\partial_{i} \delta x^{-}\right)^{2} 
  - \frac{r_{0}^{2}}{2 \nu + r_{0}^{2}} \left(\partial_{\tau} \delta x^{-}\right)^{2} \right] \,.\nonumber
\end{eqnarray}
The physical solution is found by minimizing this action.  At first order in $\epsilon$, 
the term $\partial_\tau \delta x^{-}$ vanishes since the endpoints in $\tau$ are held fixed. In addition, requiring that the remaining term proportional to $\delta r$ vanish leads to the constraint $\nu = 0$.  There is, of course, a second solution, $\nu = -\frac{4}{9} r_{0}^{2}$ but this is a maximum of the action, so we disregard it. \\

\noindent
Proceeding to the next order, to find the fluctuation spectrum, the fluctuations can be decomposed into eigenfunctions of the various differential operators in the second order action.  Fluctuations with energy $\omega$ have time dependence $\sim e^{-i \omega \tau}$ and  the spherical harmonics, $Y_l(\Omega_{3})$ are eigenfunctions of
\begin{eqnarray}
 \frac{1}{\sqrt{|g_{kl}|}} \pl_{i} (\sqrt{|g_{kl}|} g^{ij} \pl_{j} ) Y_{l} = -Q_{l} Y_{l} \,.\nonumber
\end{eqnarray}
As observed above, the deformation introduces a splitting term which, after integrating by parts, is 
$\hat{\gamma}^{2} r_{0}^{2} (\partial_{\phi_{2}} - \partial_{\phi_{3}})^{2}$.  This breaks some of the degeneracy of the spherical harmonics.  Indeed, it is clear from its form that the eigenvalue of this operator is negative semi-definite. Consequently, the spherical harmonics are diagonalized as
\begin{eqnarray}
 \left(\frac{\partial}{\partial \phi_{2}} - \frac{\partial}{\partial \phi_{3}} \right)^{2} Y_{l, \alpha} 
 = -\alpha^{2} Y_{l,\alpha} \,,
 \end{eqnarray}
As in the undeformed case, vibrations of the giant in the AdS directions $\delta x^a$ decouple from the rest of the fluctuations, and minimizing the second order action subject to the first order constraint $\nu=0$ leads to the spectrum
\begin{eqnarray} 
  \omega_a^2 = 1 + \hat{\gamma}^2 \alpha^2 + Q_l\,, 
\end{eqnarray}
which is manifestly positive definite. The coupled radial and null fluctutations, $\delta r$ and $\delta x^-$ respectively, satisfy the linear system of equations
\begin{eqnarray}
  \left[ \begin{array}{cc}
	Q_{l} - \omega^{2} + \hat{\gamma}^{2} \alpha^{2} & -2i \frac{\omega}{r_{0}} \\
	2i \frac{\omega}{r_{0}} & \frac{Q_{l} - \omega^{2}}{r_{0}^{2}}
  \end{array} \right]
  \left[ \begin{array}{c}
		\widetilde{\delta r} \\
		\widetilde{\delta x}{}^{-}
	\end{array} \right]
	 = 0\,.
\end{eqnarray}
Solving this for $\omega$ yields the spectrum
\begin{eqnarray} 
  \omega_{\pm}^{2} = Q_{l} + \left( 2 + \frac{1}{2} \hat{\gamma}^{2} \alpha^{2} \right) 
  \pm \sqrt{4 Q_{l} + \left( 2 + \frac{1}{2} \hat{\gamma}^{2} \alpha^{2} \right)^{2}} \,. 
  \label{J00-def-spectrum}
\end{eqnarray}
Here, as in the case of the undeformed giant, the fluctuation spectrum is both manifestly independent of the size of the equilibrium configuration, $r_{0}$ and {\it dependent} on the deformation parameter. 
Since the latter enters into the spectrum as a positive contribution, it is easy to see that $\omega^2$ is positive semi-definite and remains so irrespective of the deformation strength. Again, this lack of complex frequencies is a signal of the perturbative stability of these giants, exactly as was claimed in the previous section. Again, $\omega^{2}=0$ is one of the solutions of the $(\delta r,\delta x^{-})$ system when $Q_{l}=0$ reflecting the fact that the radius of the equilibrium configuration can be taken to have any value allowed by the pp-wave geometry.

\section{Comments on the dual operators}
\label{Dual-operators}
In this section we make some brief comments on the gauge theory operators dual to pp-wave giants of both the deformed and undeformed kind and leave a more detailed study of the dual giant operators and the open strings attached to them to future study \cite{Hamilton-Murugan}.
Shortly after an observation by Balasubramanian {\it et.al.} \cite{UN-giants} that the correct description of large single sphere-giant states in the dual ${\cal N}=4$ conformal field theory are of the (sub)determinant form
\begin{eqnarray}
  {\cal O}_{k} = \frac{1}{k!}\epsilon^{j_{1}\cdots j_{k}a_{1}\cdots a_{N-k}}_{i_{1}\cdots i_{k}a_{1}
  \cdots a_{N-k}} \Phi^{i_{1}}_{j_{1}}\cdots \Phi^{i_{k}}_{j_{k}}\,,
  \label{subdet}
\end{eqnarray}
rather than single-trace operators of large R-charge, it was argued by Corley, Jevicki and Ramgoolam in \cite{UN-giants} that, more generally, the operators in the $U(N)$ super Yang-Mills theory dual to giant gravitons are Schur polynomials in the Higgs fields of the SYM multiplet,
\begin{eqnarray}
  \chi_{R}\left(Z\right) = \frac{1}{n!}\sum_{\sigma\in S_{n}}\,\chi_{R}(\sigma)\,{\rm tr}\left(\sigma
  Z\right)\,.
  \label{Schur}
\end{eqnarray}
Here, $Z$ is one of three complex Higgs fields in the conformal field 
theory, $n \sim O(N)$ and 
\begin{eqnarray}
  {\rm tr}(\sigma Z) = \sum_{i_{1}\cdots i_{i_{n}}}
  Z^{i_{1}}_{i_{\sigma(1)}}Z^{i_{2}}_{i_{\sigma(2)}}\cdots Z^{i_{n}}_{i_{\sigma(n)}}\,.\nonumber
\end{eqnarray}
 The Schur polynomials are labelled by $n-$box Young diagrams and the $\chi_{R}(\sigma)$ are characters of $\sigma\in S_{n}$ in the representation $R$. The extension to the $SU(N)$ gauge theory - and consequent disentanglement of bulk and boundary degrees of freedom in giant graviton dynamics - is non-trivial and may be found in \cite{SUN-giants}. It is not difficult to see that in the totally antisymmetric representation, the Schur polynomial reduces to the subdeterminant operator (\ref{subdet}) - a single column Young diagram. Such column diagrams have a length bounded by $N$ and so encode in a very natural way the momentum cutoff observed in the dynamics of sphere giants. Similarly, a giant graviton blown up along the AdS $3-$sphere is naturally identified with the completely symmetric representation - a single row Young diagram whose length can grow without bound \cite{UN-giants,Free-fermion}. Much of the power of the machinery developed in the first of \cite{UN-giants} derives from the fact that the multipoint correlators of these Schur polynomial that encode much of the giant graviton dynamics can be 
 computed {\it exactly}. The $2-$ and $3-$point correlators, for example, are given by
\begin{eqnarray}
  \langle \chi_{R}(Z)(x_{1})\chi_{S}(\bar{Z})(x_{2})\rangle &=& \delta_{RS}f_{R}\frac{1}{(x_{1}-x_{2})^{2n_{R}}}\,,\nonumber\\
  &{}&\\
  \langle \chi_{R_{1}}(Z)(x_{1})\chi_{R_{2}}(Z)(x_{2})\chi_{R_{3}}(\bar{Z})(x_{3})\rangle 
  &=& g(R_{1},R_{2};R_{3})
  f_{R_{3}}\frac{1}
  {\prod_{i=2}^{3}(x_{1}-x_{i})^{2n_{R_{i}}}}\,,\nonumber
\end{eqnarray}
where, with the product running over the boxes of a Young diagram associated to the representation $R$, $i$ labelling the rows and $j$ the columns $f_{R} = \prod_{i,j} (N-i+j)$ and the factor $g(R_{1},R_{2};R_{3})$ is a Littlewood-Richardson coefficient - the multiplicity with which the representations $R_{3}$ appears in the tensor product of representations $R_{1}$ and $R_{2}$ \cite{FultonHarris}.\\

\noindent
The Penrose limit of the gravity theory manifests in the gauge theory as the double scaling limit in which the $U(1)$ R-charge $J$ and the number of colours 
$N$ both go to infinity in such a way that the ratio $J^{2}/N$ remains fixed \cite{BMN}. This BMN limit of the giant operators in the undeformed ${\cal N}=4$ gauge theory was first investigated in \cite{tak-tak} and we briefly summarize some of their results here. A sphere giant with large angular momentum $J$ is dual to a Schur polynomial labelled by a single column young diagram with $J$ rows. 
\begin{eqnarray}
  &\yng(1,1,1,1,1)&\nonumber\\
  R_{J}\quad \leftrightarrow\qquad &\vdots&\,\,\,\longleftarrow J {\rm\,\,boxes}\nonumber\\
  &\yng(1,1,1,1,1)&\nonumber
\end{eqnarray}
With the technology developed in the second of \cite{UN-giants} various extremal correlators obtained from considering the overlaps of combinations of Schur polynomials may be exactly computed. The normalized $3-$point correlator, for example, is
\begin{eqnarray}
  \frac{\langle \chi_{R_{J_{1}}}(Z)\chi_{R_{J_{2}}}(Z)\chi_{R_{J}}(\bar{Z})\rangle}
  {||\chi_{R_{J_{1}}}(Z)||\,||\chi_{R_{J_{2}}}(Z)||\,||\chi_{R_{J}}(Z)||} = 
  \sqrt{\frac{(N-J_{1})!\,(N-J_{2})!}{(N-J)!\,N!}}\,.
\end{eqnarray}
Using Stirling's approximation for $N!$ at large $N$ and assuming that $J^{2}/N$ is large, in the BMN limit 
\begin{eqnarray}
  \frac{\langle \chi_{R_{J_{1}}}(Z)\chi_{R_{J_{2}}}(Z)\chi_{R_{J}}(\bar{Z})\rangle}
  {||\chi_{R_{J_{1}}}(Z)||\,||\chi_{R_{J_{2}}}(Z)||\,||\chi_{R_{J}}(Z)||} \sim e^{-J_{1}J_{2}/2N}\,.
\end{eqnarray}
Consequently, there is no mixing of multi-particle states and the single-particle giant graviton state dual to the Schur polynomial (\ref{Schur}) is well defined. Similarly, the normalized overlap between a sphere giant with angular momentum $J$ and a multi-particle Kaluza-Klein state 
\begin{eqnarray}
  \frac{\langle (\chi_{1}(Z))^{J}\chi_{J}(\bar{Z})\rangle}{||(\chi_{1}(Z))^{J}||\,||\chi_{J}(Z)||} 
  = \sqrt{\frac{1}{N^{J}}\frac{N!}{J!(N-J)!}} \sim \sqrt{\frac{1}{J!}}e^{-J^{2}/2N}\,.
\end{eqnarray}
The exactly marginal Leigh-Strassler deformation of the super Yang-Mills gauge theory \cite{Leigh-Strassler}, recall, can be realised as the Moyal-like deformation,
\begin{eqnarray}
 X * Y = e^{i\pi\hat{\gamma}\left(Q^{1}_{X}Q^{2}_{Y} - Q^{1}_{Y}Q^{2}_{X}\right)}XY\,,
 \label{LM-star} 
\end{eqnarray}
of the product of fields in the ${\cal N}=4$ Lagrangian \cite{LM}. Here $(Q^{1}_{X},Q^{2}_{Y})$ are the charges of the corresponding fields under the $U(1)_{1}\times U(1)_{2}$ action. The Schur operator (\ref{Schur}) dual to the giant graviton, however, is a holomorphic polynomial in $Z$ so each factor in the product carries the same charge under $U(1)_{1} \times U(1)_{2}$. Evidently then, the operator that we expect to be dual to a giant graviton on the deformed pp-wave obtained by taking a Penrose limit about a $(J,0,0)$ geodesic, $\chi_{R}(Z)$, is blind to the $\hat{\gamma}-$deformation. This is certainly in agreement with the supergravity analysis of section 4.1. but almost trivially so. To claim agreement between the two sides of the duality, we still need to reproduce the spectrum of fluctuations about the giant configuration (\ref{J00-def-spectrum}) in the gauge theory. A much more formidable task, this is encoded in the dynamics of open strings attached to the giant operator (\ref{Schur}). Open strings in the gauge theory are given by operators of the form 
$W^{i}_{j} = \left( M_{1}M_{2}\cdots M_{n}\right)^{i}_{j}$ where the $M$'s could be Higgs fields, their covariant derivatives or even fermions. Following \cite{gauge2SUGRA}, the prescription for attaching open strings to the giant graviton consists of removing some number of $Z$'s
 that make up the giant and replacing them with the same number of words $W$. For a maximal sphere giant with a single open string attached, for example, the resulting operator takes the form (up to a normalization)
\begin{eqnarray}
  {\cal O}_{k} = \epsilon^{j_{1}\cdots j_{N}}_{i_{1}\cdots i_{N}}Z^{i_{1}}_{j_{1}}\cdots 
  Z^{i_{N-1}}_{j_{N-1}}\left(Y^{k}MY^{J-k}\right)^{i_{N}}_{j_{N}}\,, 
  \label{string-giant-operators}
\end{eqnarray}
where the string of $Y$'s build up the worldvolume of the string excitation with angular momentum $J\sim O(\sqrt{N})$ and the impurities $M$ are oscillator excitations of the string, either along the brane ($M=X$) or transverse to the giant worldvolume $(M=Z)$. In any case, hopping of the impurity along the chain introduces $\gamma-$dependent phase factors into the correlators computed from the operators (\ref{string-giant-operators}). It is the precise form of this $\gamma-$dependence that we need to match with the fluctuation spectrum about the pp-wave giant. For giants on the more general Frolov-Roiban-Tseytlin non-supersymmetric deformation of $AdS_{5}\times S^{5}$, a detailed study of such operators was carried out in \cite{DISS} where superb agreement between the gauge theory and supergravity was reported. Armed with some of the technology developed there and in \cite{DMSS}, we return to these questions shortly \cite{Hamilton-Murugan}.  

\section{Summary and outlook}
\label{Summary}
In this article we have initiated a study of giant gravitons on the pp-wave geometries arising from the Lunin-Maldacena deformation of $AdS_{5} \times S^{5}$. In particular we find two classes of giant 
graviton solutions, depending on whether the Penrose limit is taken about null geodesics carrying only one $J_{\phi_{i}}$ charge or those where all three $J_{\phi_{i}}$'s are non-zero. In the former - of which the $(J,0,0)$ pp-wave is a prototype - the giant graviton is energetically degenerate with the Kaluza-Klein point-graviton irrespective of the strength of the deformation. Like their undeformed counterparts, these spherical $D3-$branes exhibit a fluctuation spectrum that is independent of the size of the giant and absent of tachyonic modes indicating that they are stable objects in the supergravity theory. On the other hand, in section 4.2 we find that giants on the $(J,J,J)$ pp-wave suffer a deformation - not unlike that of the squashed giants of \cite{prokushkin-jabbari} - due to non-vanishing contributions from the NS $B-$field and RR $2-$form to the $D3-$brane worldvolume action. Indeed, this deformation of the $3-$sphere giant 
makes a full analytic treatment of the problem much more difficult than the $(J,0,0)$ case. Nevertheless,  we are able to make some progress in the small $\hat{\gamma}$ limit. Up to $O(\hat{\gamma}^{4})$, we find that the energy of the giant graviton is lifted with respect to the KK-graviton hinting that the giant graviton on this pp-wave wave background is energetically unstable. Indeed, for sufficiently large $\hat{\gamma}$ the second minimum of the lightcone Hamiltonian vanishes altogether. Finally, in section 6. we have concluded with some brief observations of the behaviour that we expect of the operators dual to these giant gravitons. \\ 

\noindent
Arguably, we have only just scratched the surface in the study of these configurations and many questions remain. For us, the first among these is a more complete analysis of the $(J,J,J)$ pp-wave giant. Clearly, this background deforms the $3-$sphere worldvolume of the $D3-$brane. The Lagrangian analysis carried out here has proven to be rather nontrivial. However, similar deformations have been observed in the sightly simpler context of giant gravitons in the maximally supersymmetric type IIB plane wave in the presence of a constant $B-$field in \cite{prokushkin-jabbari}. There, a fairly complete treatment of the problem was given using a Hamiltonian analysis and the emergence of a Nambu bracket structure in the worldvolume action. This then offers some hope that an analytic solution may be found here too. Secondly, this work began with the aim of determining the existence of giant graviton solutions on the Lunin-Maldacena background. To this end, we have made some progress in showing that at least the Penrose limit of this geometry supports these configurations. Nevertheless, it still remains to be seen if giant gravitons exist on $AdS_{5} \times S^{5}_{\gamma}$ and, given that these solutions exist in the completely non-supersymmetric Frolov-Roiban-Tseytlin background \cite{DISS}, it would be strange that they not exist in the ``intermediate" ${\cal N}=1$ geometry. 
There are, of course, also giants that expand in the $AdS$ directions. Since the deformation is exactly marginal, these should not be affected. However, their spectrum of fluctuations {\it will}! Exactly how, remains to be seen. Then, of course, there is the question of the operators dual to the deformed giants - some subset of the Schur polynomials that survive the $\gamma-$deformation - the open strings attached to them and the various correlators that need to be computed to match our supergravity results.\\

\noindent
It seems clear then that the study of giant gravitons on these backgrounds is an extremely fruitful area and promises to shed much needed light on a complete understanding of the gauge theory/gravity correspondence.
    
\acknowledgments 
J.M would like to dedicate this work to the memory of Thavarajoo Murugan. His presence will be greatly missed. Our gratitude is extended to Antal Jevicki, Dan Kabat, Horatiu Nastase and Amanda Weltman for valuable discussions at various stages of this work. We are especially grateful to Robert de Mello Koch for his constant support and encouragement throughout this project. J.M. was supported by an NRF (South Africa) overseas postdoctoral research fellowship at Brown University where most of this work was carried out. A.H. was supported at Columbia University by DOE grant DEFG02-92ER40699. Finally, the faculty, staff and students of the High Energy Theory group at Brown are gratefully acknowledged for all their support. 
\end{document}